\documentstyle[aaspp4,epsf]{article}

\lefthead{}
\righthead{}

\def\etal{{et al. }}
\def\Msun{M_{\sun}}
\def\cm3{{\rm ~cm}^{-3}}
\def\ltsima{$\; \buildrel < \over \sim \;$}
\def\ltsim{\lower.5ex\hbox{\ltsima}}
\def\gtsima{$\; \buildrel > \over \sim \;$}
\def\gtsim{\lower.5ex\hbox{\gtsima}}
\begin{document}
\title{A New Model for the Evolution of Light Elements in an Inhomogeneous
Galactic Halo   {\footnotesize(ApJ in press)}}

\author{Takeru Ken Suzuki$^1$ \& Yuzuru Yoshii$^2$}
\altaffiltext{1}{Department of Astronomy, School of Science,
University of Tokyo, Bunkyo-ku, Tokyo, 113-0033 Japan; Theoretical
Astrophysics Division, National
Astronomical Observatory, Mitaka, Tokyo, 181-8588 Japan;
stakeru@th.nao.ac.jp}
\altaffiltext{2}{Institute of Astronomy, School of Science,
University of Tokyo, Mitaka, Tokyo, 181-8588 Japan;
Research Center for the Early Universe, Faculty
of Science, University of Tokyo, Bunkyo-ku, Tokyo, 113-0033 Japan}

\begin{abstract}

We present predictions of the evolution of the light elements, Li, Be,
and B, in the early epochs of the Galactic halo, using a model of
supernova-induced chemical evolution
based on contributions from  supernovae (SNe) and cosmic rays (CRs), as
recently proposed by Tsujimoto \etal and Suzuki \etal.  This model has
the great 
advantage of treating various elements self-consistently, even under
inhomogeneous conditions, as might arise from stochastic star formation
processes triggered by SN explosions.  The most important prediction from our
model is that the abundances of light elements in extremely metal-poor stars
might be used as age indicators in the very early stages of an evolving halo
population, at times when the abundances of heavy elements (``metallicity'') in
most stars are dominated by local metal enrichment due to nearby SN
events, and is poorly correlated with age.

Plots of the expected frequency distribution of stars in the age {\it vs.}
elemental abundance diagram show that the best ``cosmic clock'' is the $^6$Li
abundance.  We have derived relationships among various cosmic-ray  parameters such as
energy input to CRs by SNe, the spectral shape of the CRs, and the chemical
composition in CRs, and find that we can reproduce very well
recent observations of $^6$Li, Be, and B in metal-poor stars.
Although our model is successful for certain sets of cosmic-ray parameters,
larger energy should be absorbed by energetic particles from each SN than
required to the current situation of Galactic disk.  We discuss an
alternative hypothesis of AGN activity in the early Galaxy as another possible
accelerator of CRs.

\end{abstract}


\section{Introduction}
\label{sec:intr}

Recent observations of elemental abundances in extremely metal-poor halo stars
(Mcwilliam \etal 1995, Ryan \etal 1996) uncovered evidence which suggests
that the chemical evolution of the early Galactic halo was
quite inhomogeneous.  For example, the abundances of various elements exhibit a
large scatter, even for stars having the same metallicity (as quantified by
[Fe/H]), in contradiction to the predictions of simple one-zone models
which, under the assumption that gas of the interstellar medium (ISM) is
well-mixed, suggest that overall stellar metallicity should be strongly
correlated with age.  The amount of ejected heavy elements from supernovae
(SNe) shows a strong dependence on the mass of SN progenitors, and furthermore,
this dependence is different for different elements (Woosley \& Weaver 1995;
Tsujimoto \& Shigeyama 1998).  Audouze \& Silk (1995) have further argued that
the observed abundance patterns for metal-poor stars reflect the
elemental abundance ratios of ejected elements by SNe with a discrete
range of progenitor masses.  As a result of the limited mixing of the ISM in
the early stages of the Galactic halo, the chemical compositions of these stars
are apparently providing information on SN events that occurred in their local
environment, during the first epochs of star formation.

The situation described above can be understood by a simple order of magnitude
argument as follows.  We define a volume of Galactic disk as $V_{\rm d}=2\pi
R^2 z$, where $R$ is the radius of the disk and $z$ is the half thickness. Then
it follows that $V_{\rm d} \simeq 200 \; {\rm kpc}^3$, when adopting
$R=15\;{\rm kpc}$, and $z=200\;{\rm pc}$. The SN-rate in the current disk is
about one SN per 30 yrs, and a typical SN remnant (SNR) shell can expand to
$\sim$ 100 pc before diffusing away in about 3 Myr (Tsujimoto, Shigeyama, \&
Yoshii 1999; hereafter TSY) if its expansion is not stopped by merging with
other SNR shells. Then, the swept-up volume of ISM by all the coeval SNR shells
is about $V_{\rm sw}\simeq$400 kpc$^3$, exceeding $V_{\rm d}\simeq$ 200
kpc$^3$, which implies that the SNR shells easily merge with one another.  On
the other hand, because the halo extends to 50 kpc, its total volume,
$V_{\rm h}$, is expected to be larger than that of current disk by
three to four orders of magnitude.  Provided that the total SN-rate in
the Galaxy at early epochs was less than $\sim$ 50 times of that of
today, which is acceptable according to the analysis of star formation history
by Madau \etal (1998), it follows that $V_{\rm sw}\ll V_{\rm h}$ at early epochs.
This indicates that each SNR in the halo could survive without merging with
other SNR shells. The heavy elements synthesized and ejected from a SN are
confined in its SNR shell, and do not mix easily with those originating from
other SNe, or already existing in the interstellar medium.

TSY presented a SN-induced chemical evolution model based on the assumption of
incomplete mixing of the heavy elements in the halo. In their model {\it all}
of the stars at early epochs were born in SNR shells, under the assumption that
these shells, due to their relatively high densities, are suitable sites for
star formation.  Their model explains the observed large scatter of heavy
element abundances in metal-poor halo stars very well, especially the trend of
europium abundances.  An important prediction of this model is that the
metallicity of metal-poor stars has no one-to-one correspondence with age.
Suzuki, Yoshii, \& Kajino (1999; hereafter SYK) extended this model to include
an analysis of light elements, such as $^6$Li, $^9$Be, and B
($^{10}$B+$^{11}$B), which are mainly produced by reactions involving Galactic
cosmic rays (GCRs), and demonstrated that this model also reproduces the
observed trends of $^9$Be and B data. SYK proposed a new scenario that GCRs
originate directly from acceleration of the SN ejecta, and also from
acceleration of particles in the ISM swept up by SNR. SYK pointed out that the
one might expect that the abundance of light elements, $^6$LiBeB, in metal-poor
stars should exhibit a much better correlation to their age, in sharp contrast
to the situation for their heavy element abundances.

There are (at least) two significant issues concerning the production of light
elements in the early Galaxy which must be confronted.  The first is that the
observed linear trend of Be and B with metallicity contradicts with that
derived from the usual spallation processes of energetic protons or
$\alpha$-particles impinging on CNO elements in the ISM.  The
production rate of the light elements appears {\it constant} with respect to
metallicity, when we take the observations at face value, while the usual
spallation processes predict that the production rate should be {\it
proportional} to metallicity, so that a quadratic trend of Be and B with
metallicity is expected.  As one possible solution, Duncan \etal (1992)
and Yoshii, Kajino, \& Ryan (1997)
proposed an ``inverse'' spallation process applied in the early Galaxy, with
energetic CNO nuclei originating from freshly synthesized SN-ejecta impinging
on protons and $\alpha$-particles in ISM producing Be and B.  The production
rate predicted from these inverse processes is almost constant as a
function of metallicity, because SNe of metal-poor progenitors synthesize a
comparable amount of CNO elements to metal-rich SNe.  The condition for these
processes to work is that synthesized CNO elements in SNe should be efficiently
injected into the acceleration region around the SNR shock.  However, based on
a reanalysis of elemental abundances of present-day GCRs and an examination of
acceleration process thought to be provided by the SNR shock, Meyer \& Ellison
(1999) and Ellison \& Meyer (1999) have recently argued that most GCRs were
accelerated out of the ISM or circumstellar material instead of SN ejecta,
although they suggested that, owing to the lack of CNO in the ISM at early
epochs, energetic CNO out of fresh SN ejecta might still play a role in light
element production.  As a compromise solution, SYK claimed that if $\sim 2$\%
of GCRs originate from SN-ejecta, in agreement with the estimate by Meyer
\& Ellison (1999) and Ellison \& Meyer (1999), the linear trend of BeB with
metallicity can be achieved for stars with [Fe/H]$<-1.5$.  Fields \& Olive
(1999) argued that oxygen, instead of iron, should represent the
``metallicity,'' because BeB elements are mainly produced by spallation of
oxygen in the early stages of the Galaxy.  They claimed that a quadratic BeB
trend with O results, when using the increasing trend of [O/Fe] inferred from
the observations of Israelian \etal (1998) and  Boesgaard \etal (1999) for the
stars of extremely low metallicity.  However, this trend of oxygen relative to
iron is still controversial, and also differs from the trend of other
$\alpha$-elements such as Mg (Fields \etal 1999 also mentioned this problem).

The second issue concerns the shortage of available energy to accelerate GCRs
in order to produce the observed level of abundances of light elements. Ramaty
\etal (1997) pointed out the importance of the ``energy budget''
of GCRs when investigating the evolution of light elements, and they concluded
that GCRs should be more metal-rich than the ambient matter to account for the
observed Be abundance in metal-poor stars.  The energetics of the production of
light elements in SNRs has been intensively studied by
Parizot \& Drury (1999a,b; hereafter PDa,b).  They also concluded that energy
input by individual SNe is about one order of magnitude lower than that
required to reproduce the observations.  To solve this discrepancy, the
a superbubble model was suggested by Higdon \etal (1998) as the source
of metal-rich energetic particles (EPs).  Since the material in superbubbles
becomes metal rich due to collective explosions of SNe, enough CNO nuclei can
be accelerated by successive SNe, with no need for the artificial selective
acceleration of CNO nuclei.  From extensive calculations, Parizot \& Drury
(1999c; hereafter PDc) concluded that superbubbles are capable of producing
enough of the light elements to satisfy the observed level.  It should be
noted, however, that the superbubble model cannot simultaneously explain
the large scatter of heavy elements in metal-poor stars (see more discussion of
this point in \S \ref{sec:sbmd}). In this sense the second problem still
persists.

In this paper we formulate the model of SN-induced chemical evolution in \S
\ref{mdlgce}, and the model of EPs in \S \ref{sec:gcr} and \S \ref{sec:leps}.
We argue that an acceptable relation among three unknown parameters (input
energy to EPs per SN, the spectral shape of EPs, and composition of EPs) can
be found that reproduces the observed abundance of light elements (\S
\ref{sec:ccpr}).  The results of our model are presented in \S \ref{sec:rslt}.
We compare the predicted evolution of light elements with recent
observations in \S \ref{sec:cmpobs}, and we predict the evolution of
the composition of GCRs in the early stage of the Galactic halo in \S
\ref{sec:cmpcr}.  In \S \ref{sec:tmr} we discuss the feasibility of using
the abundances of each of the light elements as age indicators for old
metal-poor stars.  Other related topics are discussed in \S \ref{sec:dscs}.
Our model is compared with an alternative model of superbubbles as a
source of light elements in \S \ref {sec:sbmd} in terms of energetics of 
GCRs, and the possibility of considering AGNs as another accelerator of
EPs is discussed in \S \ref{sec:agn}. 

\section{Evolution of Elements}
\label{mdlgce}

In this section we present formulae which describe the chemical evolution
in an inhomogeneous early Galaxy, circumstances that are expected to arise from
the stochastic nature of star-forming processes induced by SN explosions.
Star-forming processes are assumed to be confined in separate clouds of mass
$M_{\rm c}$ which make up the entire halo at early epochs.  We set $M_{\rm
c}=10^{7}\Msun$ throughout all the calculations in this paper, because
the results 
of the calculation do not depend crucially on the value of this parameter.  The
evolution of a cloud starts at time $t=0$ when a certain fraction, $x_{\rm
III}$, of the cloud turns into metal-free Pop III stars with an
initial mass function, $\phi(m)$, having a Salpeter index of $-1.35$ with upper
and lower mass limits of $m_u=50\Msun$ and $m_l=0.05\Msun$, respectively.  In
our calculations, $x_{\rm III}$ is set to be $10^{-4}$, which satisfies the
condition that more than one Pop III star explodes as a Type II SNe (SNe II)
to trigger successive star-forming processes (TSY).  Chemical evolution
characterized 
by star formation processes is triggered by SN explosions.  Since the velocity
of ejected matter in SN explosions exceeds the sound velocity there, a shock
front is formed and swept-up ISM material will be accumulated behind the front
to form a dense shell.  Although the temperature of the shell is quite high at
first, the shell gradually cools as it loses energy (mainly by radiative
losses), and it will eventually form cool and dense fragments. Some of these
fragments might become seeds of new stars.  All stars of subsequent generations
are assumed to form in these shells behind the radiative shock front.  The mass
fraction, $\epsilon$, of each shell that turns into stars, is taken as a
constant. Here, we adopt $\epsilon=4.3\times10^{-3}$, which gives the best fit
to the observed [Fe/H] distribution function for various values of $x_{\rm
III}<10^{-2}$ (TSY).  Then, the star formation rate (SFR) at time $t$ is given
by

\begin{equation}
\label{eqn:sfr}
\hspace{-0.03cm}\displaystyle{
\dot{M_\ast}(t)}=\int_{\max(m_t{\rm ,}\,m_{{\rm SN},\,l})}^{m_u}
\hspace{-1.8cm}dm{\epsilon}M_{\rm sh}(m,t)\frac{\phi(m)}{m}
\dot{M_\ast}(t-{\tau}(m)){\rm ,}
\end{equation}
where ${\tau}(m)$ denotes the lifetime of a star with mass $m$, and $m_t$ 
is the stellar mass for which ${\tau}(m)=t$. A lower mass limit 
for stars that explode as SNe is taken to be $m_{{\rm SN},\,l}=10\Msun$.
The mass of the shell is given by

\begin{equation}
M_{\rm sh}(m,t)=M_{\rm ej}(m)+M_{\rm
sw}(m,t){\rm ,}
\end{equation}
where $M_{\rm ej}(m)$ is the mass of the SN ejecta,
and $M_{\rm sw}(m,t)$ is the mass of the gas swept up by the SNR,
given by $M_{\rm sw}(m,t)=6.5{\times}10^4{\Msun}({E_{\rm
SN}}/{10^{51}{\rm erg}})^{0.97}$ as a function of explosion energy,
$E_{\rm SN}$, per SN (Shigeyama \& Tsujimoto 1998; TSY).

Using the SFR in eq.(\ref{eqn:sfr}), the mass of gas, $M_{\rm g}$, changes
with time according to the star formation and the gas ejection from 
stellar mass loss and SN explosions (TSY):  

\begin{equation}
\label{eqn:chevg}
\hspace{-0.03cm}\displaystyle{dM_{\rm g} \over dt}=-{\dot M}_\ast
(t)+\int_{\max ({m_t{\rm ,}\, m_l})}^{m_{u}}\hspace{-1.2cm}
dmM_{\rm ej}(m){\phi(m) \over m}{\dot M}_\ast ({t-\tau (m)}){\rm ,}
\end{equation}

The abundance of the $j$-th heavy element in the gas, $Z_{j,{\rm g}}(t)$, 
changes with time according to the formula (TSY):

\begin{displaymath}
\hspace{-1.5cm}\displaystyle
{d(Z_{j,{\rm g}}M_{\rm g}) \over dt}
=-\int_{\max ({m_t, m_{{\rm SN}, \,l}})}^{m_{u}}
\hspace{-1.8cm}dm
Z_{j, \ast} ({m {\rm ,}\, t})\epsilon M_{\rm sh}({m {\rm ,}\, t}) 
{\phi (m)\over m }
+\int_{\max ({m_t{\rm ,}\, m_l})}^{m_{u}}\hspace{-1.2cm}
dm(M_{\rm ej}(m)- \sum_iM_{Z_i}(m)){\phi(m)\over
m}
\end{displaymath}
\begin{displaymath}
\hspace{-0.15cm}
\times\int_{\max({m_{t-\tau(m)}{\rm ,}\,m_{{\rm SN},\,l}})}^{m_u} 
\hspace{-2.2cm}dm^\prime Z_{j, \ast} ({m^\prime
{\rm ,}\, t-\tau (m)})\epsilon M_{\rm sh}({m^\prime {\rm ,}\, t-\tau
(m)}){\phi ({m^\prime }) \over m^\prime }{\dot M}_\ast ({t-\tau(m)-\tau
({m^\prime })}) 
\end{displaymath}
\begin{equation}
\label{eqn:chevm}
+\int_{\max ({m_t{\rm ,}\, m_l})}^{m_{u}}\hspace{-1.2cm}
dmM_{Z_j}(m){\phi (m)\over m}\int_{\max ({m_{t-\tau (m)}{\rm ,}\,
m_{{\rm SN}, l}})}^{m_{u}}\hspace{-2.7cm}
dm^\prime \epsilon M_{\rm sh}({m^\prime
{\rm ,}\, t-\tau(m)}){\phi ({m^\prime }) \over m^\prime }{\dot
M}_\ast({t-\tau(m)-\tau({m^\prime })}),
\end{equation}
where $M_{Z_j}(m)$ is the mass of synthesized $j$-th heavy element
ejected from a star with mass $m$, and $Z_{j, \ast} ({m {\rm ,}\, t})$ is
the stellar abundance of the $j$-th element born at time $t$ from a SNR
shell with progenitor mass $m$.
The second term denotes the $j$-th
ejected element that has survived through stellar evolution after being
trapped in stars, and the third term is the ejected element
which is newly synthesized through the stellar evolution and SN
explosion.   
Since stars are formed in the SNR shell that contains {\it both}
SN ejecta and swept-up ISM, the stellar metallicity can be written as
(TSY): 

\begin{equation}
\label{eqn:shabcno}
\hspace{-1cm}\displaystyle
Z_{j, \ast}(m,t)=\frac{M_{Z_j}(m)+Z_{j, \ast}(m,t-{\tau}(m))
(M_{\rm ej}(m)-\sum_iM_{Z_i}(m))
+Z_{j, {\rm g}}(t)M_{\rm sw}(m,t)}{M_{\rm sh}(m,t)} \;\;.
\end{equation}
 
As light elements are easily burned by the stellar nuclear processes at 
temperatures of a few $10^6$K, we can assume that these elements, once taken up
by stars, are quickly destroyed.  Some of light elements are produced by GCRs
{\it globally} propagating in the cloud, and others by the EPs confined in each
SNR that will be thermalized without escaping from the SNR. The local EPs will
produce the light elements in the same SNR by collisional reactions and the
amount of the produced light element $L$ in each SNR, $M_{Z_{L,lcr}}(m,t)$,
depends on the amount of CNO ejected from each SN having a variety of
progenitor mass $m$.  These locally produced light elements will be modeled
later in \S \ref{sec:leps}.  Then, the abundance of the $L$-th element in the
gas, $Z_{L,{\rm g}}(t)$, changes as follows:

\begin{displaymath}
\hspace{-1cm}\displaystyle
{\frac{d(Z_{L, {\rm g}}M_{\rm g})}{dt}}=-\int_{\max 
(m_t,m_{{\rm SN}, l})}^{m_u}
\hspace{-1.8cm}dmZ_{L, \ast}(m,t){\epsilon}M_{\rm sh}(m,t)
\frac{\phi{(m)}}{m}
\dot{M_\star}(t-{\tau}(m))
\end{displaymath}
\begin{displaymath}
\hspace{-1cm}+\sum_{i=p\alpha,j={\rm CNO}}(\langle\sigma_{ij}^L
F_i\rangle Z_{j, {\rm g}}(t)(A_L/A_j)+\langle\sigma_{ji}^LF_j\rangle X_i(t)
(A_L/A_i)){M_{\rm g}(t)}
\end{displaymath}
\begin{displaymath}
+\int_{\max ({m_t{\rm ,}\, m_{{\rm SN,}l}})}^{m_{u}}\hspace{-1.2cm}
dmM_{Z_{L,{\rm lcr}}}(m,t){\phi (m)\over m}
\int_{\max ({m_{t-\tau (m)}{\rm ,}\,
m_{{\rm SN}, l}})}^{m_{u}}\hspace{-2.7cm}
dm^\prime \epsilon M_{\rm sh}({m^\prime
{\rm ,}\, t-\tau(m)}){\phi ({m^\prime }) \over m^\prime }{\dot
M}_\ast({t-\tau(m)-\tau({m^\prime })})
\end{displaymath}
\begin{equation}
\label{eqn:chevle}
\hspace{-0.1cm}+\int_{\max ({m_t{\rm ,}\, m_{{\rm SN,}l}})}^{m_{u}}
\hspace{-1.2cm}
dmM_{Z_{L,\nu}}(m){\phi (m)\over m}\int_{\max ({m_{t-\tau (m)}{\rm ,}\,
m_{{\rm SN}, l}})}^{m_{u}}\hspace{-2.7cm}
dm^\prime \epsilon M_{\rm sh}({m^\prime
{\rm ,}\, t-\tau(m)}){\phi ({m^\prime }) \over m^\prime }{\dot
M}_\ast({t-\tau(m)-\tau({m^\prime })}){\rm ,}
\end{equation}
with
\begin{equation}
\langle\sigma_{ij}^LF_i\rangle\equiv\int_{E_{\rm th}}^{\infty}
\sigma_{ij}^L(E)F_i(E, t)S_L(E)dE{\rm ,}
\end{equation}
where $A_i$ is the mass number of the $i$-th element and $X_i$ is the abundance
of hydrogen or helium. $\sigma_{ij}^L(E)$ is the cross section for the process
of the GCR projectile $i$ impinging on the ISM target $j$ to produce the $L$-th
element.  $S_L(E)$ gives the retention fraction of $L$-th product that can
survive to be thermalized in the ISM, and $F_{i}(E, t)$ is the time-dependent
flux of GCR projectile $i$, which is modeled in \S \ref{sec:gcr}.  The second
term in eq.(\ref{eqn:chevle}) represents the production by global GCRs, and the
third term production by EPs confined in each SNR. $M_{Z_{L,\nu}}(m)$ in the
last term represents the mass of the ejected $L$-th element synthesized by the
neutrino process just after SN explosions (Woosley \etal 1990), which is only
significant for the production of $^{7}$Li and $^{11}$B (Vangioni-Flam \etal
1996).  The yield tables of Woosley \& Weaver (1995) are used for the
$\nu$-process, but the absolute values are decreased by a factor of 5 in order
to reproduce the observed  $^{10}$B/$^{11}$B ratio (Vangioni-Flam \etal 1996,
1998) in our calculation.  The stellar abundance of the $L$-th element is equal
to that in the SNR shell and is given by

\begin{displaymath}
\hspace{-1cm}\displaystyle
Z_{L,\ast}(m,t)=[\sum_{i=p\alpha,j={\rm CNO}}
(\langle\sigma_{ij}^LF_i\rangle Z_{j,{\rm g}}(t)
(A_L/A_j)+\langle\sigma_{ji}^LF_j\rangle X_i(t)
(A_L/A_i)) M_{\rm sh}(m,t){\Delta}T
\end{displaymath}
\begin{equation}
\label{eqn:shable}
+M_{Z_{L,{\rm lcr}}}(m)+M_{Z_{L,\nu}}(m)+Z_{L,{\rm g}}(t)M_{\rm sw}(m,t)]
/M_{\rm sh}(m,t)\; ,
\end{equation}
where $\Delta T=3\times 10^6\: {\rm yrs}$ is a typical diffusion time of a SNR
shell (TSY).  The first term in the numerator represents the mass of the $L$-th
element produced during $\Delta T$ by {\it global} GCRs originating from all
the SNe that explode at time $t$. The second and third terms are the masses of
the $L$-th element produced by {\it local} EPs and that by the neutrino
process, respectively, and the last term represents the mass of the $L$-th
element included in the swept-up material.

\section{Models of Energetic Particles}

In our scenario all the EPs originate from SN explosions. Those particles
absorb the energy of the explosions by being scattered back and forth across
the shock front (first-order Fermi acceleration). Some of them are scattered
far away, upstream of the blast wave's (forward) shock, and
will propagate as GCRs, while some of the EPs are trapped inside the SNR and
will produce the light elements there by inelastic collisional reactions
(PDa,b). These two types of EPs give different results, from the viewpoint of
the spatial inhomogeneity of the abundance of light elements.  It is expected
that {\it global} GCRs will produce these elements uniformly in the entire
cloud at the same epoch by the spallation of CNO elements in the ISM, while
{\it local} EPs confined in each SNR will produce these elements in a way that
their abundance depends on the amount of CNO elements ejected from different
progenitor masses of SNe.  In the next two subsections the models of these two
types of EPs are presented.

\subsection{Global Cosmic Rays}
\label{sec:gcr}

Some of the energetic particles (accelerated as described above) can escape
from the SNR, and will propagate far outside the SNR shell.  These EPs, or
GCRs, are expected to be distributed much more uniformly over the patchy
structure of the early ISM. Although most of these particles are expected to
originate from acceleration of the swept-up ISM by the forward shock, some of
them come from SN-ejecta injected in the acceleration region behind the forward
shock by various mechanisms, which are considered in detail below.  SN ejecta,
if condensed into grains, can be injected more effectively into the
acceleration region because the mass-to-charge ratio for grains is higher, and
it follows that the the Larmor radius is larger (Lingenfelter \etal 1998).

Mass loss from progenitor stars before the SN-explosions take place makes it
possible for the forward shock to easily accelerate the previously-ejected
metal-rich material surrounding the central stars.  Progenitor stars of SNeII
lose their mass over their entire lifetimes, mainly by radiation-driven stellar
winds (e.g.  Chiosi \& Maeder 1986). Therefore, at the time of the
SN-explosion, a circumstellar envelope composed of ejected material during the
pre-SN-explosion era is thought to exist around the central star.  Such
circumstellar envelopes have been observed in several SNRs (Plait \etal 1995
for SN1987A; Benetti \etal 1998 for SN1994aj; Chu \etal 1999 for SN1978K). In
particular, the mass ejected by the explosion of SN1994aj is estimated to be
3-5$\Msun$, while the mass of progenitor star in the main sequence phase is
thought to have been 8--20$\Msun$, which indicates that about half or more of the
initial mass of the progenitor star has been lost during the pre-SN-explosion
era.  Observations of the abundance ratios of CNO elements in the ring around
SN1987A show that it consists of the material synthesized in the stellar
interior (Panagia \etal 1996).  This fact implies that even the products
synthesized in the deep stellar interior can be dredged up to the stellar
surface by convection (e.g. Maeder 1987; Heger \etal 2000), and ejected by
stellar winds to form the circumstellar envelope before the SN explosion.
Although the above considerations are based on observations of present-day SNR,
even very metal-poor old stars are expected to contain significant amounts of
heavy elements in their outer envelopes during later evolutionary stages, from
this ``self-pollution'' process.  These dredged-up heavy elements would play a
crucial role to drive stellar winds as major opacity sources.  As a result,
more metal-rich circumstellar envelopes than the ambient ISM are expected to be
formed, and the forward shock created after the SN-explosions can accelerate
such metal-rich materials originating from stellar nucleosynthesis, as well as
the metal-poor ISM, with similar efficiency.

Owing to the processes described above, these EPs accelerated by the
forward shock are supposed to be a mixture of metal-rich stellar and SN ejecta
(we hereafter call ``SN ejecta'' for simplicity) and metal-poor ISM. SYK
proposed a new model that takes into account these two origins of
GCRs using a free parameter, $f_{\rm cr}$, as defined below. We employ the
same parameterization.  When the momentum spectrum, which is expected from
shock acceleration theory (Blandford, \& Ostriker 1978; Blandford, \& Ostriker
1980), is used, the source spectrum of global GCRs in units of ${\rm
particles{\:} s^{-1}g^{-1}(MeV/A)^{-1}}$ at time $t$ can be written as

\begin{equation}
\label{eqn:crfx}
q_i(E,t){\propto}\frac{(E+E_0)}{[E(E+2E_0)]^{\frac{\gamma+1}{2}}}\int_{\max
({m_t{\rm ,}\,
m_{{\rm SN},l}})}^{m_{u}}\hspace{-1.2cm}dm\{M_{Z_i}(m)+Z_{i,{\rm g}}(t)f_{\rm cr}M_{\rm sw}(m,t)\}
{\phi(m) \over A_im}{\dot M}_\ast ({t-\tau (m)}),
\end{equation} 
where $E_0$ is the rest mass energy of a nucleon $E_0=930\;{\rm MeV/A}$,
$\gamma={(r+2)}/{(r-1)}$ is the spectral index which is
related to the compression ratio (the velocity
difference), $r$, of the shock (Blandford, \& Ostriker 1978; Blandford,
\& Ostriker 1980), $M_{Z_i}(m)$ is the mass of the $i$-th heavy 
element synthesized and ejected from an SN with progenitor mass $m$,
and $f_{\rm cr}$ is the fractional mass of the gas in the shell swept up while
the SN explosion is able to accelerate ISM particles. It should be noted that
$f_{\rm cr}$ {\it determines} the elemental composition of GCRs.

The total flux of the source is normalized by the input energy of SNe.  If we
define $E_{\rm gcr}$ as the energy used to accelerate particles into GCRs per
SN, and $\dot{N}_{\rm SN}(t)$ as the SN-rate at time $t$ in a given cloud (which
can be calculated from the formula presented in \S \ref{mdlgce}), then
$q_i(E,t)$ is related to $E_{\rm gcr}$ and $\dot{N_{\rm SN}}$ as follows:

\begin{equation}
\label{eqn:crnm}
M_{\rm g}(t)\sum_{i}\int^{E_{\rm max}}_{E_{\rm min}}dE\:E\:q_i(E,t)=E_{\rm
gcr}\dot{N}_{\rm SN}(t)\; ,  
\end{equation}
where $E_{\rm max}=10^{14}\;{\rm eV/A}$ is the highest energy
achieved by SN explosions and $E_{\rm min}$ is the
low-energy cutoff. Provided $2<\gamma<3$, particles with energy $E\ll E_0$ make
little contribution to the total integrated energy carried by the bulk of GCRs,
so we here adopt $E_{\rm min}=0.1\;{\rm MeV/A}$.

These EPs will propagate as GCRs in a cloud and interact with the ambient
medium.  The propagation of GCRs is taken into account by using the leaky-box
model (Meneguzzi \etal 1971). When ``grammage'', $X\;({\rm g\:
cm^{-2}})$, is used 
as an independent variable, the transport equation for the energy spectrum of
the flux of the $i$-th element, $F_i(E,t)$, is expressed as

\begin{equation}
\label{eqn:eptp}
\frac{\partial F_i(E,t)}{\partial X}=
q_i(E,t)+\frac{\partial}
{ \partial E}[\omega_i(E){F_i(E,t)}]-\frac{F_i(E,t)}{\Lambda_{\rm
esc}}-\frac{F_i(E,t)}{\Lambda_{{\rm 
n},i}}\; ,
\end{equation}
where $q_i(E,t)$ is the source spectrum (taken to be the same as in
eqs.(\ref{eqn:crfx}) and (\ref{eqn:crnm})), and ${\omega}_i(E)$ is the 
ionization energy losses in ${\rm MeV/A (g\:cm^{-2}) ^{-1}}$ through a
hydrogen-helium plasma with $X_{\rm H}$=0.75 and $X_{\rm He}$=0.25, as
tabulated in Northcliffe 
\& Schilling (1971). $\Lambda_{\rm esc}$ is the loss length in  ${\rm g\:
cm^{-2}}$ due to escape from a given region, and $\Lambda_{{\rm n},i}$ is that
against nuclear destruction, given by (Malaney \& Butler 1993)

\begin{equation}
\label{eq:ptlngt}
\Lambda_{{\rm n},i}=\frac{M_p+(n_{\alpha}/n_p)M_{\alpha}}{\sigma_{pi}
+(n_{\alpha}/n_p)\sigma_{{\alpha}i}}\; ,
\end{equation}
where $M_p$ and $M_{\alpha}$ are the masses of protons and $\alpha$-particle,
respectively, $n_{\alpha}/n_p$ is the ratio of $\alpha$ to proton number
density in the ISM, and $\sigma_{pi}$ and $\sigma_{{\alpha}i}$ are the total
cross sections of nuclear reactions of particle $i$ interacting with protons
and $\alpha$-particles, respectively.  Using the tabulated cross sections (Read
\& Viola 1984), one can determine $\Lambda_{{\rm n},p}\simeq 200{\rm g\:
cm^{-2}}$ for protons, and $\Lambda_{\rm n, \alpha CNO}\simeq 20{\rm g\:
cm^{-2}}$ for $\alpha$ and CNO particles, for the energy range of
50$\sim$500$({\rm MeV/A})^{-1}$ where the light elements are dominantly
produced.  We set the escape length to be ${\Lambda_{\rm esc}}=100{\rm g\:
cm^{-2}}$. The choice of this value does not change the production rate of
light elements, provided that ${\Lambda_{\rm esc}}>\Lambda_{\rm n,\alpha CNO}
(\simeq 20{\rm g\: cm^{-2}})$.  This is because (1) at early epochs almost all
the BeB are produced by spallation reactions of GCR CNO, and $^{6}$Li is
produced by the $\alpha + \alpha$ fusion reaction, and (2) the loss from
escape becomes negligible compared to that against the nuclear destruction of
$\alpha$ and CNO particles under those conditions.

Equation (\ref{eqn:eptp}) does not take into account spatial
inhomogeneity of GCRs, which means that GCRs are assumed to have
one-zone features in each cloud. The time scale of GCR
transport across the cloud, $\tau_{\rm trsp, c}$, is estimated as
 
\begin{eqnarray}
\label{eq:difcr}
\tau_{\rm trsp, c} &\simeq & \frac{R_{\rm c}^2}{2D_{\rm GCR}} \nonumber \\
&\simeq& 1\;{\rm Myr}\left(\frac{R_{\rm c}}{1\;{\rm
kpc}} \right)^2 \left(\frac{D_{\rm GCR}}{10^{29}{\rm cm^2\:s^{-1}}}\right)^{-1}
\end{eqnarray}
where $R_{\rm c}\sim (0.1-1)\;{\rm kpc}$ is a typical cloud size, and $D_{\rm
GCR}\simeq 10^{29}{\rm cm^2\:s^{-1}}$ is the diffusion coefficient of
GCRs in the Galactic halo obtained from observed elemental compositions of EPs
(e.g., III \S 3 in Berezinsli$\breve{\rm i}$ \etal 1990) and electron component
of EPs (e.g., V \S 12 in Berezinsli$\breve{\rm i}$
\etal 1990).  The time scale of the cloud evolution, $\tau_{\rm
evol,c}$($\simeq 20$ Myr), is characterized by the lifetimes of stars with $m
\simeq 10\Msun$ -- it is their SN explosions that induce the formation of new
stars.  Provided that the properties of GCR transport in the early Galaxy are
similar to the current epoch, it follows that $\tau_{\rm evol,c} \gg
\tau_{\rm trsp, c} $, which indicates that GCRs propagate throughout the cloud
much faster than the cloud evolution.  Therefore, we can assume that GCRs are
distributed uniformly in each cloud, and the use of eq.(\ref{eqn:eptp}) is
justified.  In our scenario, star-forming processes are
confined to occur in separate clouds with mass $M_{\rm c}\;(\sim 10^7\Msun)$,
making up an entire Galactic halo ($\sim 10^{11}\Msun$). Thus, the early halo
consists of $\sim 10^4$ such clouds.  Some clouds start their
evolution with SN explosions of first-generation stars (metal-free Pop III
stars) earlier than others, so that clouds have their own evolutionary
histories arising from different SFRs. Some of the GCRs originating from a
given SN in a cloud can leak out of it and reach different clouds.  While we
incorporate the contribution of these GCRs coming from other clouds just by
increasing the escape length, $\Lambda_{\rm esc}$, we do not perform a more
precise modeling of different histories of SFR or SN-rates in different clouds.
However, the dominant component of GCRs in each cloud is obviously composed of
GCRs originating from SNe that exploded in the same cloud, because the flux of
GCRs coming from other clouds is decreased considerably, being inversely
proportional to the square of distance (under the assumption of isotropic
propagation).  Moreover, the average history of star formation in the Galactic
halo is constrained to reproduce the observed [Fe/H] distribution of metal-poor
stars (\S \ref{mdlgce}), so that typical clouds are expected to have, more or
less, a similar star formation history to that considered in our model.  Thus,
our model, which assumes that the flux of GCRs is in proportion to the SN-rate
in the cloud, can account for the expected evolution of typical clouds in the
Galactic halo.

\subsection{Local Energetic Particles and Light Element Production in
the SNR}
\label{sec:leps}

It is commonly assumed that at least some EPs are trapped in SNR shells by the
diffusion barriers around the shock front, even after getting enough energy to
produce the light elements (e.g., VII \S 4 in Berezinsli$\breve{\rm i}$ \etal
1990).  For the purpose of establishing a more sophisticated model to
investigate global evolutionary trends of the light elements, it is
indispensable to take into account the production of light elements by these
EPs trapped inside each SNR, in addition to the production by GCRs propagating
globally outside the SNRs.  EPs in the SNR, and their production of light
elements during the Sedov-Taylor phase, have been intensively studied by PDab.
Based on their model, we consider the region inside the SNR as a viable
production site for light elements.  Some of the trapped EPs will produce
light elements in the SNR, lose their energy, and never escape.
In our model, we define $E_{\rm lcr}$ (the subscript indicates ``local cosmic
rays'') as the amount of energy per SN absorbed by the particles which receive
enough energy to produce the light elements, but subsequently lose that energy
and become thermalized in the same SNR.  The aim in this subsection is to
derive the yield term, $M_{Z_{L,{\rm lcr}}}(m)$ (eqs. (\ref{eqn:chevle}) \&
(\ref{eqn:shable})) of the light elements contributed from nuclear processes
involving these confined EPs, as correlated with the mass $m$ of the
progenitor to the SN.

Since the typical evolution timescale (${\lesssim} 10^6$ yrs) of the SNR
is much 
shorter than the typical timescale ($\sim 10^9$ yrs) for the chemical
evolution of the clouds in the Galactic halo, one has to adopt a much
shorter timescale when considering production of light elements in the evolving
SNR, which is usually approximated as an instantaneous event in the context of
the chemical evolution of the clouds.  We define $t'$ as the time elapsed after
{\it each} SN event, which should be distinguished from the time $t$ elapsed
after the formation of first stars in the cloud, so that $t'$ is used as an
independent variable to describe the phenomena occurring in the evolving SNR.

According to standard SNR theory, the forward shock survives until the
SNR shell loses its identity, while the reverse shock exists only at very the
beginning of the Sedov-Taylor phase (Truelove \& McKee 1999).  Therefore,
since no energy is input to EPs after the reverse shock disappears, the
particles accelerated by the reverse shock more easily lose their energy and
are trapped in the SNR, as compared with those accelerated by the forward
shock.  For simplicity, we assume that these particles which are finally
thermalized in the SNR are those accelerated by the reverse shock, and the
acceleration occurs only at the very beginning of the Sedov-Taylor phase.  We
here define $t_{\rm ST}$ as the time when the Sedov-Taylor phase starts, 
corresponding to the time when the swept-up mass is equal to the ejected mass,
$M_{\rm ej}(m)$. The time, $t_{\rm ST}$, is on order of 1000 years,
assuming typical 
densities, ejected mass, and released kinetic energy of a SN (PDa).  Since the
reverse shock is formed in the ejecta (Truelove \& McKee 1999), all of the
particles in this process are assumed to come from SN ejecta. Thus, the source
spectrum for mass $m$ of SN progenitor in units of ${\rm particles{\:}
s^{-1}g^{-1}(MeV/A)^{-1}}$ is expressed as

\begin{equation}
Q_i(E,m,t'){\propto}\frac{(E+E_0)}{[E(E+2E_0)]^{\frac{\gamma+1}{2}}}
M_{Z_i}(m)\delta(t'-t_{\rm ST}),
\end{equation}
where $Q_i(E,m,t')$ is normalized by the condition : 
\begin{equation}
M_{\rm ej}(m)\sum_{i}\int^{\infty}_{0}dt'\int^{E_{max}}_{E_{min}}
dE\:E\:Q_i(E,m,t')=E_{\rm lcr}\; ,
\end{equation}
where the notation is the same as in the previous section, and
the transport of these particles is treated as follows:

\begin{equation}
\frac{\partial N_i(E,m,t')}{\partial t'}=
Q_i(E,m,t')\rho+\frac{\partial}{ \partial E}[\{\dot{E}_i(E)+\dot{E}_{\rm
ad}(E,t')\}{N_i}]-\frac{N_i}{\tau_{{\rm n},i}}\; .
\end{equation} 
This equation is similar to eq.(\ref{eqn:eptp}), though $t'$ is used as an
independent variable instead of the grammage, $X$.  Generally, $X$ is related
to time $t'$ by $X=\rho v t'$, where $\rho$ is the density of the ambient matter,
and $v$ is the velocity of the EPs.  $N_i(E,m,t')$ is the energy spectrum of
the $i$-th element formed at time $t'$ in units of  ${\rm particles{\:}} {\rm
s^{-1}cm^{-3}}({\rm MeV/A})^{-1}$, which is related to the flux by
$F_i=N_{i}v_{i}$, where $v_{i}$ is the velocity of $i$-th EP.  $\dot{E}_{\rm
ad}(E)$ is the adiabatic loss rate, and $\dot{E}_i(E)$ is the ionization loss
rate in units of ${\rm (MeV/A)\:s^{-1}}$. $\dot{E}_i(E)$ can be calculated from
$\omega_i(E)$ in eq.(\ref{eqn:eptp}) by $\dot{E}_i(E)= \omega_i(E)\rho v_i$.
$\tau_{{\rm n},i}$ is the time scale of nuclear destruction of the $i$-th
element, which is also related to $\Lambda_{{\rm n},i}$ in eq.(\ref{eqn:eptp})
via $\tau_{{\rm n},i}=\Lambda_{{\rm n},i} \rho v$. Here, we do not explicitly
add the term for the escape of particles out of the SNR, because this effect
can be incorporated phenomenologically by changing the values of $E_{\rm gcr}$
and $E_{\rm lcr}$.  Following PDab, the adiabatic loss rate is given by

\begin{equation}
\dot{E}_{\rm ad}(E,t')=-\frac{3}{10}\frac{E}{t'}(\frac{E+2E_0}{E+E_0}). 
\end{equation}

We would like to address how much the change of $\rho$ changes the production
of the light elements. When the ionization and nuclear destruction dominate the
loss, different $\rho$'s would give the same total production, because the
production rate is linearly proportional to $\rho$ in the same way as the loss
timescale, so that the $\rho$-dependencies are cancelled out in multiplying the
production rate and the loss time to give the total production.  On the other
hand, the adiabatic loss time only weakly depends on $\rho$, so higher $\rho$
gives rise to more target nuclei to produce light elements.  The
$\rho$-dependence of the production of light elements has been considered also
by PDab, and our model confirms their basic result.  As far as light-element
production is concerned, the effect of changing $\rho$ is similar to ``tuning''
a value of $ E_{\rm lcr}$.  Therefore, in this paper, we use a fixed value of
$\rho=1\times 10^{-23}\:{\rm g\: cm^{-3}}$ $(\sim 5\:{\rm atom\;cm^{-3}})$.
 
The target nuclei in spallation processes involved with the production of the
light elements are thought to be swept up in the SNR shell, with
mass given by

\begin{equation}
M_{\rm SNR}(m,t')=M_{\rm ej}(m)+\frac{4}{3}\pi R^3(t')\rho \; {\rm ,}
\end{equation}
where $R(t')\propto t'^{2/5}$ is taken from the Sedov-Taylor similarity
solution (Taylor 1950; Sedov 1959).  Then, the total mass of the $L$-th
element, $M_{Z_{L,{\rm lcr}}}(m,t)$, produced by a SN of progenitor mass $m$
can be calculated by integrating the production rate of light elements over
$t'$ through the SNR phase :

\begin{displaymath}
M_{Z_{L,{\rm lcr}}}(m,t)=\int^{t_{\rm max}}_{t_{\rm ST}}\hspace{-0.4cm}
dt'\hspace{-0.4cm}\sum_{i=p\alpha,\\j={\rm CNO}}\hspace{-0.3cm}[\langle\sigma_{ij}^L
F_i\rangle(m,t')Z_{j, {\rm SNR}}(m,t,t')(A_L/A_j)
\end{displaymath}
\begin{equation}
+\langle\sigma_{ji}^LF_j\rangle(m,t') X_{i,\rm SNR}(m,t,t')
(A_L/A_i)]{M_{\rm SNR}(m,t')}{\rm ,}
\end{equation}
with
\begin{equation}
\langle\sigma_{ij}^LF_i\rangle(m,t')\equiv\int_{E_{\rm th}}^{\infty}
\sigma_{ij}^L(E)F_i(E,m,t')S_L(E)dE{\rm ,}
\end{equation}
where $t_{\rm max}$ is the time at which EPs are no longer
capable of producing light elements (due to energy loss), $X_{i,\rm SNR}(m,
t')$ is the abundance of either hydrogen or helium in the SNR of progenitor
mass $m$ at time $t'$, and $Z_{j, {\rm SNR}}(m,t')$ is the $j$-th element
abundance, given by

\begin{equation}
Z_{j,{\rm SNR}}(m,t,t')=\frac{M_{Z_j}(m)+Z_{j, \ast}(m,t-{\tau}(m))
(M_{\rm ej}(m)-\sum_iM_{Z_i}(m))
+Z_{j, {\rm g}}(t)\frac{4}{3}\pi R^3(t')\rho}
{M_{\rm SNR}(m,t')} \;\; .
\end{equation}

\subsection{Constraints On Elemental Production}
\label{sec:ccpr}

The model presented in the previous sections has three free parameters that
character the global GCRs -- the spectral index, $\gamma$, of EPs at the
source, the parameter $f_{\rm cr}$, which quantifies the proportion
of GCRs originating from SN ejecta and the swept-up ISM, and the energy,
$E_{\rm gcr}$ per SN, that accelerates the GCRs.  In order to take into
account local processes involving EPs in the SNR shell, we need another
parameter, $E_{\rm lcr}$ per SN, absorbed by the EPs which are 
thermalized in the SNR shell.  In the real situation, $\gamma$ has different
values for different processes, and is time-dependent as the physical state of
the shock changes.  However, here we employ a single value of $\gamma$ as an
average over different processes and phases.

In this subsection we constrain the values of these free parameters based on
$^{6}$Li observations reported by Smith \etal (1998), Cayrel \etal
(1999), and Nissen \etal (2000) and $^{9}$Be by Boesgaard
et al.  (1999).  $^{6}$Li is produced by $\alpha+\alpha$ fusion and the
spallation of CNO elements, whereas $^{9}$Be is produced exclusively by the
spallation of CNO.  Both $^{6}$Li and $^{9}$Be abundances have been measured
in two metal-poor stars, HD 84937 and BD +26$^{\rm o}$ 3578, yielding a
$^{6}$Li to $^{9}$Be ratio of 50$\sim$100, though the ratio of spallation cross
sections to produce $^{6}$Li and $^{9}$Be is as small as 5. This implies that
$^{6}$Li is mainly produced by $\alpha+\alpha$ fusion in the early Galaxy.
Since the helium in the universe was mostly produced by Big-Bang
nucleosynthesis (BBN), and distributed globally in the halo, the abundance of
$\alpha$-particles is expected to be almost constant in either the ISM or GCRs.
So the production rate of $^{6}$Li depends little on the composition of heavier
nuclei in GCRs, but rather, is determined by the flux of EPs in the relevant
energy range, that is, the spectral shape, $\gamma$, and energy, $E_{\rm gcr}$,
necessary to accelerate GCRs.
 
\begin{figure}[p]
\epsfxsize=13cm
\epsfysize=18cm
\epsfbox{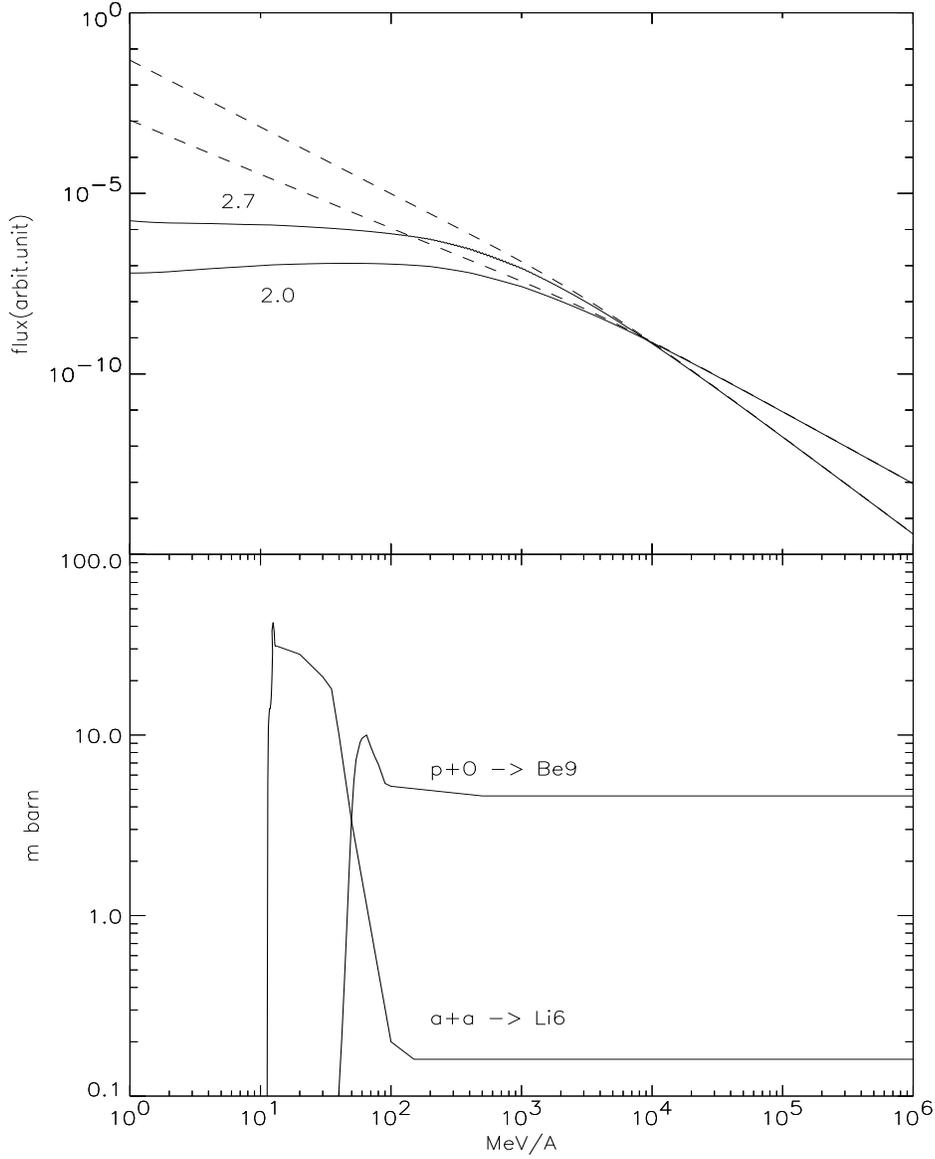}
\caption{{\it top panel} : Spectrum of GCRs for the slope index of 
$\gamma=2.0$ and 2.7. Solid and dashed lines show the transported and
source spectrum, respectively. 
{\it bottom panel} : Production cross section for $^6$Li ($\alpha +
\alpha$ fusion) and Be (O+p spallation) as a function of energy
per nucleon.}
\end{figure}

The top panel of Fig. 1 shows the transported flux (solid lines) and the
source flux (dashed 
lines) of GCRs for two cases, $\gamma=2.0$, and $\gamma= 2.7$.  The total input
energy to EPs at the source is taken to be the same for both $\gamma$'s, so
that the flux of EPs in the low energy region of $\sim$100MeV/A for the softer
spectrum case ($\gamma=2.7$) is one order of magnitude higher than for the
other case.  The cross section for producing $^{6}$Li by $\alpha$ + $\alpha$
fusion strongly depends on the energy of incident EPs, and has a maximum
value in the low-energy region of 10-100MeV/A (the bottom panel of
Fig. 1). Therefore, the 
production rate of $^{6}$Li is very sensitive to the spectral shape of EPs --
the softer spectrum obviously results in more $^6$Li production (See
also Vangioni-Flam \etal 1999).  Thus, we are
able to constrain the relation between $\gamma$ and $E_{\rm gcr}$ from the
observed $^{6}$Li to Fe ratio. At present there are only three reported
detections of $^{6}$Li.  For simplicity, we here assume that Fe abundances in
these three stars are quite similar to those in the well-mixed gas. The shaded
area in Fig.2 denotes the allowed relation between $\gamma$ and $E_{\rm gcr}$,
which accounts for the observed data of log($^{6}$Li/H)$\simeq (-11.0
\sim -11.5) \pm 0.4$ at [Fe/H]$=(-2.2 \sim -2.3) \pm 0.2$ presently
obtained.  The reported detections have rather large associated errors,
which results in a large allowed region of parameter space.  
However, we should caution that the Fe
abundance of metal-poor stars has no one-to-one correspondence with age
(TSY, SYK, and see more discussion in \S \ref{sec:tmr}).  Therefore, it is
necessary to constrain the $\gamma$-$E_{\rm gcr}$ relation by comparing the
$^6$Li data {\it directly} with the theoretical frequency distribution
of the stellar [Fe/H]-log($^{6}$Li/H) plane (see, e.g., Suzuki \etal 1999 for a
likelihood analysis of Li ($^{6}$Li+$^{7}$Li)).  Precise determination of
$^6$Li abundance in many metal-poor stars is highly desirable, as it will
provide much stronger constraints on the adopted parameters of EPs in the early
Galaxy.

$^{9}$Be is produced only by the spallation of CNO nuclei, which implies that
the production rate depends on the CNO abundance in both the ISM and GCRs.  At
early epochs, because of the low abundance of CNO nuclei in the ISM, most of
the $^{9}$Be is produced by GCR CNO from SN ejecta spalling with $p \alpha$ in
the ISM (see \S \ref{sec:cmpcr}).  This implies that the $^9$Be production rate
should depend linearly on the amount of ejected CNO from SNe. Since our model
considers no metallicity dependence of CNO yields from SNe, the $^9$Be
production rate is proportional to the total amount of EPs coming from SN
ejecta.  We define a proportionality coefficient

\begin{equation}
\label{eqn:ejrt}
\theta= \frac{E_{\rm gcr,ej}}{E_{\rm gcr}}+ \frac{\alpha_{\rm
ad}{E_{\rm lcr}}}{E_{\rm gcr}}{\rm ,}
\end{equation}
where ${E_{\rm gcr,ej}}$ is the energy used to accelerate GCRs
originating from SN-ejecta.  The first term is the energy ratio of GCRs
coming from SN ejecta relative to the total GCRs, and is expressed in
terms of the mass ratio as 

\begin{equation}
\label{eqn:ejrte}
\frac{E_{\rm gcr,ej}}{E_{\rm gcr}}=\frac{\overline{M_{\rm ej}}}{f_{\rm
cr}M_{\rm sw}+\overline{M_{\rm ej}}}\;{\rm ,}
\end{equation}
where ${\overline{M_{\rm ej}}}$ is the IMF-weighted average of ejected mass
from SNe.  The second term in eq. (\ref{eqn:ejrt}) represents the
contribution from the local EPs (\S \ref{sec:leps}), denoted by the energy
ratio of local to global EPs.  The reduction factor, $\alpha_{\rm ad}$, is
introduced in order to take into account the effect that the $^{9}$Be
production is reduced due to the adiabatic loss of EPs in the SNR.
Thus, as in the case of $^{6}$Li, the existent 
$^9$Be data can be used to derive the relation between $\gamma$ and $E_{\rm
gcr}$ for a given value of $\theta$.  In contrast to $^{6}$Li,
there are enough stars with measured $^9$Be abundances (Boesgaard \etal 1999).
Thus, it is possible to derive the $\gamma$-$E_{\rm gcr}$ relation by comparing
the $^9$Be data with the theoretical frequency distribution of stars in the
[Fe/H]-log($^{9}$Be/H) plane (see Fig.3).  Solid lines in Fig.2 show the
allowed $\gamma$-$E_{\rm gcr}$ relation for various values of $\theta$.  Since
the spallation cross sections of CNO producing $^9$Be have, more or less, a
constant value (only the spallation cross section of oxygen is shown for
reference in the bottom panel of Fig.1), compared to those for the $\alpha+\alpha$ fusion
producing $^6$Li, it is apparent that the allowed $\gamma$-$E_{\rm gcr}$
relation only weakly depends on the spectral index $\gamma$.

\begin{figure}[p]
\begin{flushleft}
\epsfxsize=17cm
\epsfysize=13cm
\epsfbox{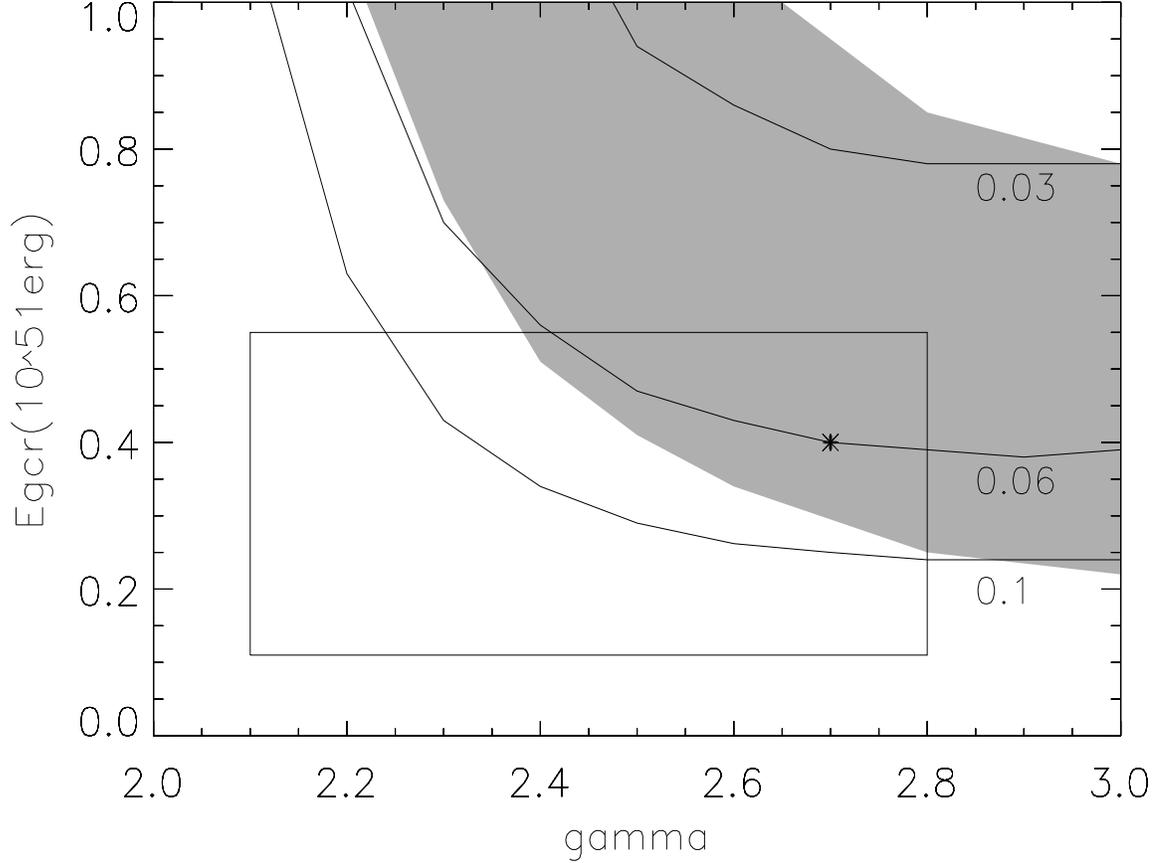}
\end{flushleft}
\caption{Relation between spectral index, $\gamma$, of EPs
and energy, $E_{\rm gcr}$, used to accelerate GCRs per SN that accounts for
the observation of $^6$Li (Smith \etal 1998, Cayrel \etal 1999, and
Nissen \etal 2000) and $^9$Be (Boesgaard \etal 1999) in metal-poor stars.
Shaded area shows an allowed region from $^6$Li data, and solid
lines denotes the allowed relations from $^9$Be observation for various
values of $\theta$. The vertical length of the box corresponds to a
range of uncertainty from numerical simulations of shock acceleration in 
the SNR by Berezhko \& V${\rm \ddot{o}}$lk (1997), and the horizontal
length of the box corresponds to the uncertainty in observation of
high-energy particles. The asterisk * indicates the value adopted as our 
standard.}
\end{figure}

Three free parameters -- $\gamma$, $E_{\rm gcr}$, and $\theta$, are constrained
by the data for $^{6}$Li and $^{9}$Be, still leaving one degree of freedom.
Let us now return to previous work where the constraints on $E_{\rm
gcr}$ and $\gamma$ are provided.  The clearest constraint on the energy is that
$E_{\rm gcr}$ cannot exceed the total kinetic energy of a SN explosion, $E_{\rm
SN}$.  Another, rather strong constraint, from recent simulations of shock
acceleration is that $10\%\sim 50\%$ of the explosion energy is absorbed by EPs
(Berezhko \& V${\rm \ddot{o}}$lk 1997) in the SNR.  Therefore, it is necessary
to precisely determine $E_{\rm SN}$ in order to constrain $E_{\rm gcr}$. X-ray
observations (Hughes, Hayashi, \& Koyama 1998) of seven SNRs in the Large
Magellanic Cloud show that the explosion energy of each SN exhibits a range
of about one order of magnitude from $5\times 10 ^{50}$erg to $6\times 10
^{51}$erg when the observed data are fit by a model using the Sedov-Taylor
similarity solution. According to these authors, three of the seven SNRs seem
to have exploded within pre-existing cavities, thus the application of Sedov
model fits is in some doubt.  The average value for remaining four SNRs,
which are thought to have exploded in the usual circumstances, is $E_{\rm
SN}=(1.1\pm 0.5)\times 10^{51}$erg, although these authors pointed out that
this value is a lower limit, since they assumed spherically symmetric SNRs,
which results in the lowest estimate of $E_{\rm SN}$. The case for their result
of $E_{\rm SN}=1.1\times 10^{51}$erg with the numerical simulation (Berezhko \&
V${\rm \ddot{o}}$lk 1997) is indicated in Fig. 2.

The spectral index can also be constrained from observations of high-energy
particles on the Earth's surface.  The observed proton and helium spectra with
energy above a few GeV/A is best fit with $\gamma_{\rm obs} = 2.7 \sim 2.8$
(Burnett et al. 1983; Webber 1987; Webber et al. 1987), and the CNO energy 
spectrum is almost identical (e.g., Engelmann et al. 1985). In order to explain
the energy-dependent secondary/primary ratio of B/C, the present escape
length, $\Lambda_{\rm esc}$, of GCRs relies on the 
rigidity dependence, or equivalently, on the energy of incident EPs, as far as
elements with the same mass-to-charge ratio are considered.  According to
Garcia-Munoz \etal (1987), in the higher energy range of
$E \gtsim$1GeV/A, it is  
expected that  $\Lambda_{\rm esc} \propto E^{-0.6}$, while in the lower energy
range the exact energy dependence of $\Lambda_{\rm esc}$ is difficult to obtain
because of the strong influence of solar modulation.  From the observed
value of $\gamma_{\rm obs}$ in the higher energy range, a source spectral index
is inferred to be $\gamma_{\rm source}=\gamma_{\rm obs}-0.6=2.1\sim2.2$, which
is flatter than the results, $\gamma_{\rm source}=2.3\; {\rm and}\; 2.36$,
obtained from the data for elemental ratios of primary nuclei in 0.1-100
GeV/A by Webber \etal (1992) and Lukasiak \etal (1994), respectively.  If the
effect of re-acceleration of EPs in the ISM is taken into account, it is
allowed that $\Lambda_{\rm esc}$ has lower rigidity dependence, leading to
$\Lambda_{\rm esc} \propto E^{-1/3}$ over the entire energy range (Seo \&
Ptuskin 1994; Strong \& Moskalenko 1998).  An inferred source spectrum depends
on the strength of re-acceleration.  For example, it is reported that
$\gamma_{\rm source}=2.4$ (Seo, \& Ptuskin 1994), and $\gamma_{\rm
source}=2.25$ (Strong \& Moskalenko 1998).  Numerical simulations of the shock
acceleration in SNRs also results in a variety of source spectra, with
$\gamma_{\rm source}=2.2-2.5$, depending on the physical state of the SNR
(V$\ddot{\rm o}$lk, Zank, \& Zank 1988).  Thus, it seems that a decisive
constraint on the spectral index has not yet been obtained, due to the
complicated influence of various loss and acceleration processes involving EPs.
Moreover, these arguments are largely based on the observations 
in the current Galactic disk, and the situation in the early Galactic halo may
be different.  Therefore, we presently place a conservative constraint that the
source spectral index may fall between the flattest case ($\gamma =2.1$)  and
the observed one ($\gamma=2.8$) -- this is the allowed range shown in
Fig.2.

In this paper we adopt, as a compromise, the set of the parameter values
$E_{\rm gcr}=4.0\times10^{50}{\rm erg}$, $\gamma=2.7$, and $\theta=0.06$, which
satisfy all the above constraints.  According to eq.(\ref{eqn:ejrt}), $\theta$
is factored into two terms involving global GCRs and local EPs -- the
reduction factor in the EP term is taken to be $\alpha_{\rm ad}\sim 0.25$ for
$\gamma=2.7$ and $\rho=1\times 10^{-23}\:{\rm g\:cm^{-3}}$.  Since all of the
parameters in $\theta$ are quite uncertain, their values should be constrained
from the theory of shock acceleration in SNRs.  In the proceeding sections,
we chose ${E_{\rm lcr}}/{E_{\rm gcr}}=0.1$ and $f_{\rm cr}=0.007$, which
correspond to the situation that 3.5$\%$ of the total GCRs come from
the SN ejecta, with the rest coming from the swept-up ISM.  In \S
\ref{sec:tmr}, we discuss the case for different values of ${E_{\rm
lcr}}/{E_{\rm gcr}}$.

\section{Results}
\label{sec:rslt}

\subsection{Comparison with Observations}
\label{sec:cmpobs}

We now show the results of our model predictions.  The initial abundances of
heavy elements are set to be zero, but those of the light elements are set
equal to the primordial abundances of log($^6${Li}/H)=$-14.5$,
log(Be/H)=$-17.9$,  and log(B/H)=$-16.9$, based on the standard BBN calculation
of Thomas et al.  (1994).  The chemical evolution of the star-forming cloud
described in \S \ref{mdlgce} starts from the epoch of Pop III star formation,
and terminates at 0.6 Gyr, when SNRs sweep up all the material of the clouds
(TSY).  At that time, the metallicity of the cloud reaches [Fe/H] $\sim -1.5$.
The parameters for EPs are adjusted to reproduce the observational data (\S
\ref{sec:ccpr}).  Our results for the predicted frequency distribution of
long-lived stars (m$<1\Msun$) in the [Fe/H] {\it vs.} log($X_{L}$/H) plane
are compared with the $^6$Li data from Smith \etal (1998), Cayrel \etal
(1999), and Nissen \etal (2000) in the top panel of Fig.3, the Be
data from Boesgaard et al. (1999) in the middle of Fig.3, and the B
($^{10}$B+$^{11}$B) data 
from Duncan et al. (1997) and Primas et al. (1999) in the bottom of
Fig.3. In order to 
directly compare the model predictions with observations, frequency
distributions are convolved with observational errors assumed to have Gaussian
dispersions $\sigma = 0.15 $ dex for Be, B, and Fe, and $\sigma = 0.3 $ dex for
$^{6}$Li.  We define a probability density of finding one halo star within a
unit area of $\Delta$[Fe/H]=0.1$\times\Delta$log($X_L$/H) =0.1, normalized
to unity 
when integrated over the entire area.  Two contour lines shown in the figures
are, from the inside out, of constant probability density of $10^{-3}$ and
$10^{-5}$, respectively.

\begin{figure}[p]
\epsfxsize=13cm
\epsfysize=18cm
\epsfbox{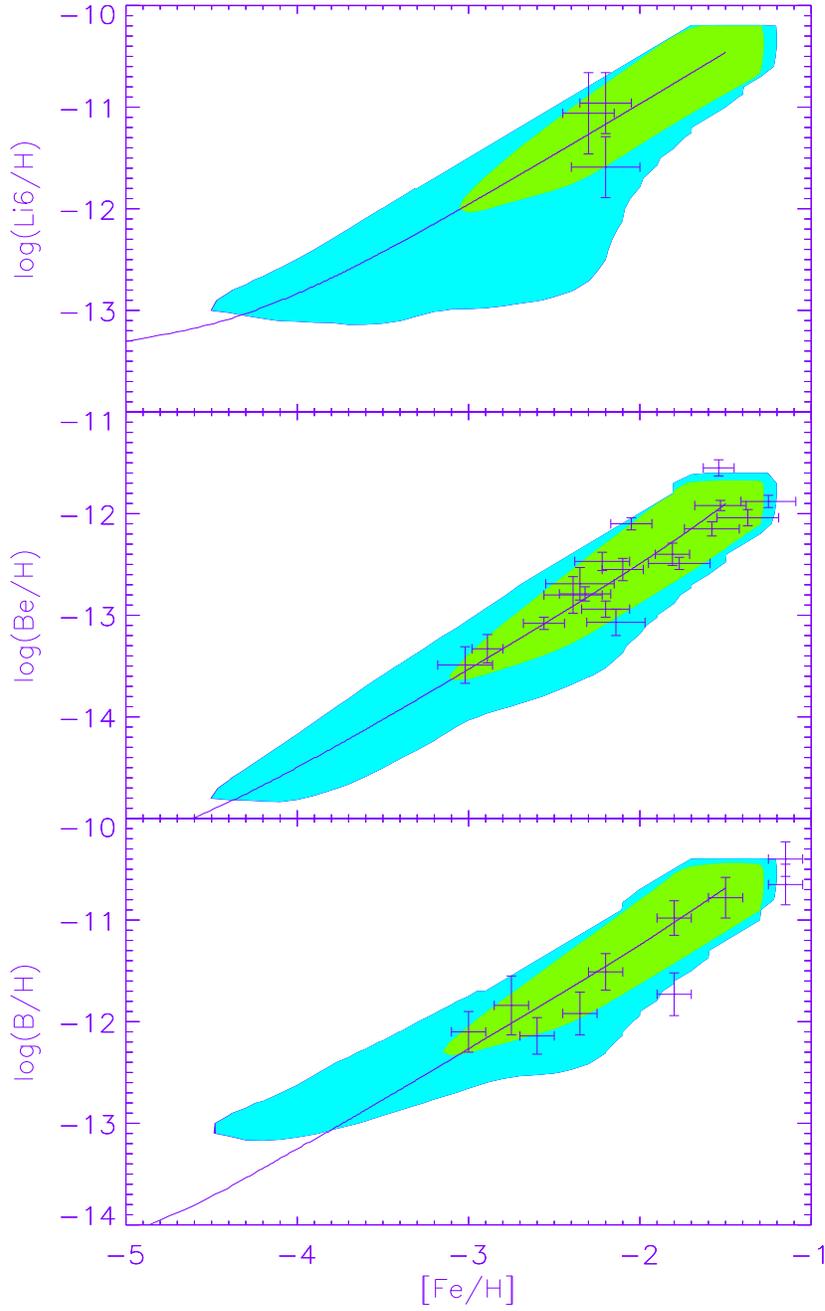}
\caption{Predicted frequency distributions of long-lived stars in the
[Fe/H]-log($X_L$/H) plane ($X_L$=$^{6}${Li} for {\it top panel},
$X_L$=$^{9}$Be for {\it middle panel}, and $X_L={\rm B}$ for {\it
bottom panel}), 
convolved with Gaussian having 
$\sigma = 0.15 $ dex for Be, B, and Fe and $\sigma = 0.3 $ dex for
$^{6}$Li.
Two contour lines, from the inside to the outside, correspond to those of
constant probability density $10^{-3}$, and
$10^{-5}$ in unit area of
$\Delta$[Fe/H]=0.1$\times\Delta$log($X_L$/H)=0.1.  The 
solid line shows the [Fe/H]-log($X_L$/H) relation in the gas. The crosses
represent the data with observational errors 
taken from Smith \etal (1998), Cayrel \etal (1999), and Nissen \etal
(2000) for $^{6}$Li,  Boesgaard et al. (1999) for 
Be, and Primas \etal (1999) and Duncan \etal (1997) for B.}
\end{figure}

Our model predictions are in good agreement with the observed $^6$Li, Be, and B
abundances.  In particular, the distributions of the Be and B data in the
[Fe/H] {\it vs} log($X_{L}$/H) plane appears to be consistent with the area
of constant probability density of $10^{-3}$. This implies that if the number
of stars with measured Be and B abundances are increased by a factor of 100,
they will fill in the area of constant probability density of $10^{-5}$.  The
observed linear trend of Be and B with Fe in a range of [Fe/H]$>-3$ is well
reproduced, because most of Be and B arise from the spallation of GCR CNO
from SN ejecta, due to the lack of CNO in the ISM at early epochs (\S
\ref{sec:cmpcr}).
  
\subsection{Composition of Cosmic Rays}
\label{sec:cmpcr}

The top panel of Fig. 4 shows the evolution of C+O abundance in GCRs and
the ISM as a 
function of the time elapsed after the formation of metal-free Pop III stars.
We adopt $f_{\rm cr}=0.007$, which corresponds to the case that 3.5$\%$ of GCRs
originate from SN ejecta.  The C+O abundance in GCRs is quite high compared to
that in the ISM (but only a factor of 3--4 lower than the solar value
$\sim$ 0.02), and is almost constant during the entire halo phase, though
slightly increasing toward the end of the phase due to the increase of C and O
in the swept-up ISM which become GCRs.  In order to achieve such a high 
C + O abundance in the GCRs, the acceleration of stellar and SN ejecta
by the forward shock of SN explosion, discussed in \S
\ref{sec:gcr}, plays an important role. On the other hand,
the C+O abundance in 
the ISM is much lower than that in GCRs, even though it is an increasing
function of time due to the chemical evolution of the halo.  Accordingly, the
inverse process of GCR CNO + ISM p$\alpha$ is quite important for the
production of light elements in the early phases of the Galaxy.

\begin{figure}[p]
\epsfxsize=13cm
\epsfysize=18cm
\epsfbox{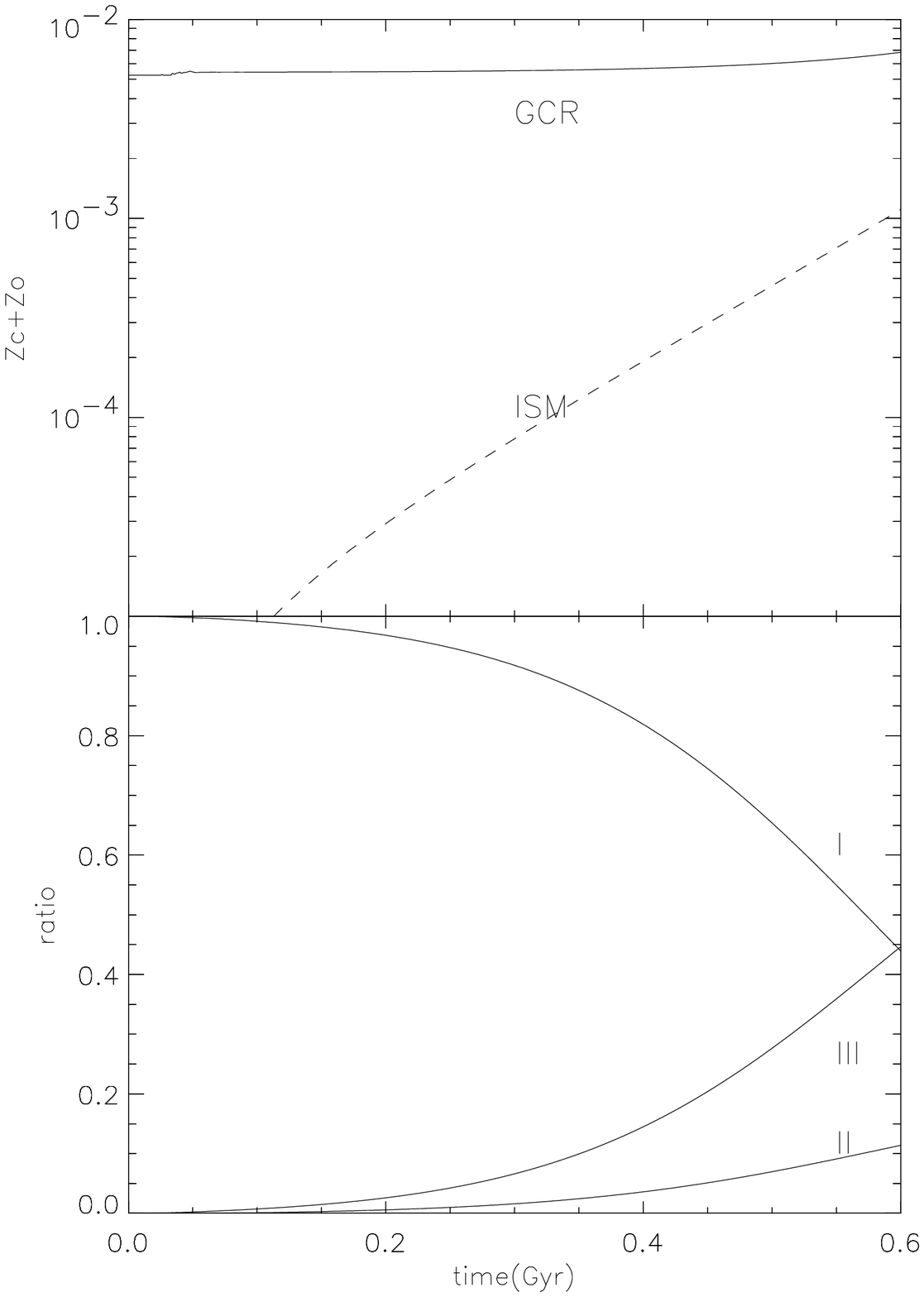}
\caption{ {\it top panel} : Evolution of C+O abundance in GCRs
(solid line), and in the gas (dashed line).
 {\it bottom panel} : Each contribution of processes I, II ,and III
(see text) to the total production of $^9$Be as a function of time.}
\end{figure}

Spallation of CNO coming directly from SN ejecta results in a linear relation
between the abundances of spallation products (the light elements) and heavy
elements originating from SNe, while spallation of CNO which were once
thermalized in the ISM results in a quadratic relation.  Therefore, the
reactions of GCR CNO from SN ejecta + ISM p$\alpha$ (process I) bring about a
linear trend, whereas GCR CNO from the ISM + ISM p$\alpha$ (process II) and GCR
p$\alpha$ + ISM CNO (process III) produce a quadratic trend.  The
bottom panel of Fig. 4 shows
the contributions of these three processes in the $^9$Be production as a
function of time.  In the beginning, almost 100$\%$ of $^{9}$Be is produced by
process I, and the contributions by the other two processes gradually increase
with increasing CNO abundances in the ISM.  At the end stage of an evolving
halo, the contribution of process III becomes similar to that of the process I.
In our model, during the entire halo phase ([Fe/H]$<-1.5$), process I dominates
over the other two, leading to a linear dependence of $^{9}$Be abundance on
metallicity, in good agreement with observations (\S \ref{sec:cmpobs}).
                 
\subsection{Stellar Age {\it vs.} Elemental Abundance Relation}
\label{sec:tmr}

In past work the abundance of heavy elements observed in a given star has been
regarded as indicative of the time at which that star was formed.  Although
this is expected from  simple one-zone models of chemical evolution, its basic
assumption of a well-mixed gas in a closed nucleogeneric zone should be
re-examined when considering the very early stages of an evolving halo.  The
abundances of heavy elements in metal-poor stars reflect those of synthesized
elements by individual SNe that have just exploded near the site of formation
of such stars.  However, SYK recently pointed out that the abundances of the
light elements, except for $^{7}$Li, in very metal-poor stars can still be used
as age indicators because these elements are mainly produced by the reactions
involving GCRs that propagate globally.  In this subsection we examine the
feasibility of using various light element abundances as age indicators.

\begin{figure}[p]
\epsfbox{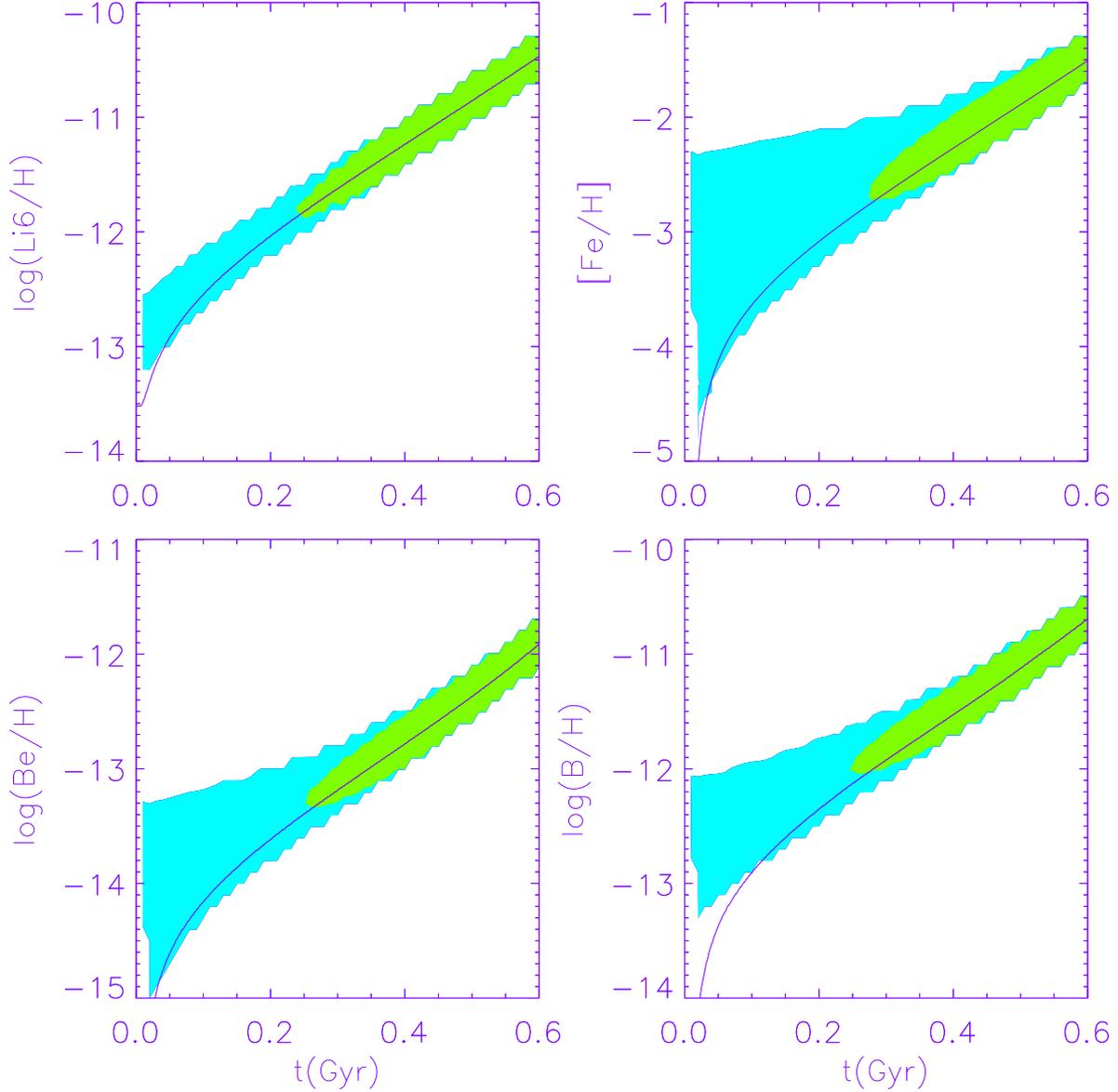}
\caption{ Predicted frequency distribution of long-lived stars in the 
$t$-log($X_i$/H) planes as a function of time,
convolved with Gaussian having 
$\sigma = 0.15 $ dex for $X_i$=Be, B, and Fe and $\sigma = 0.3 $ dex for
$^{6}$Li. 
Time, $t$, is defined as that elapsed after the formation of metal-free
Pop III stars. Two contour lines, from
the inside to the outside, correspond to those of constant probability
density $10^{-3}$, and $10^{-5}$ in unit area of $\Delta t=10({\rm
Myr})\times\Delta$log($X_i$/H)=0.1.
Solid lines represent the evolution of abundances of the elements in the 
gas.}
\end{figure}

Figure 5 shows the abundances of Fe, B, Be, and $^{6}$Li as a function of time
elapsed after the formation of Pop III stars. The contours in this figure
show the frequency distribution of long-lived stars with $m<1\Msun$,
born at time $t$, and the solid line represents the elemental abundances in the
ISM.  We define a probability density of finding one halo star within a unit
area of $\Delta$log($X_i$/H)=0.1$\times \Delta t=10$ Myr, normalized to unity
when integrated over the entire area.  The two contour lines shown are, from
the inside out, of constant probability density $10^{-3}$ and $10^{-5}$.

Generally speaking, the narrower distribution of stars along the solid line
indicates a better correlation between stellar age and elemental abundance.  In
this respect, the predicted $t$-Fe correlation is considerably poorer at
earlier epochs. A better correlation is tenable only at [Fe/H]$>-2$, where
Fe can be used as an average age indicator.  The ideal element for a cosmic
clock is $^{6}$Li, for which a superior correlation is realized at all times
from the beginning.  This results because $^{6}$Li is mainly produced by the
fusion of $\alpha$-particles, which are the BBN products, and are distributed
globally throughout the entire halo.  The $t$-Be correlation is marginally
acceptable for use as a cosmic clock, because it is produced by the
spallation of CNO in GCRs which propagate globally.  B may also be used -- it
is a better clock than Fe, but worse than Be, because a significant fraction of
$^{11}$B isotope is produced in the SNR shell by the $\nu$-process of SNe II
in addition to the spallation of CNO in GCRs.

In our model, heavy elements observed in metal-poor stars originate from
SN ejecta and the swept-up ISM, and all light elements originate mainly from
the spallation of CNO elements, as well as the $\alpha + \alpha$ fusion
producing Li.  Table 1 summarizes the sources that produce each of the elements
considered in this paper.

\begin{table}[h]
\caption{Origin of Elements}
\label{tab:oe} 
\begin{center}
\begin{tabular}{cc}
\tableline
\tableline
element & source\\ \tableline
Fe & SNe II \\
B  & SNe II($\nu$-process), GCRs \& local EPs (spallation)\\
Be & GCRs \& local EPs (spallation) \\
$^{6}$Li & GCRs \& local EPs (mainly fusion) \\
\tableline
\end{tabular}
\end{center}
\end{table}

The heavy and light elements originating from these different processes are
mixed in the SNR shell from which stars are formed with chemical composition
according to eqs.(\ref{eqn:shabcno}) and (\ref{eqn:shable}).  The numerators in
these equations are separated into two parts -- one is the ISM term, which
reflects the homogeneous nature of chemical composition in the ISM for either
heavy or light elements

\begin{equation}
M_{z_j,{\rm ISM}}(t)=Z_{j, {\rm g}}(t)M_{\rm sw}{\rm ,}
\end{equation}
and the other is the SN-shell term, which reflects the inhomogeneous nature
of the generation of heavy elements,

\begin{equation}
M_{z_j,{\rm sh}}(m,t) = M_{Z_j}(m)+Z_{j,
\ast}(m,t-{\tau}(m))\nonumber 
(M_{\rm ej}(m)-\sum_iM_{Z_i}(m)),
\end{equation}
and for light elements, 

\begin{eqnarray}
M_{z_L,{\rm sh}}(m,t) &=& \sum_{i=p\alpha,j={\rm CNO}}
(\langle\sigma_{ij}^LF_i\rangle Z_{j,{\rm g}}(t)
(A_L/A_j) \nonumber \\
& &+\langle\sigma_{ji}^LF_j\rangle X_i(t)
(A_L/A_i)) M_{\rm sh}(m){\Delta}T \nonumber \\
& &+M_{Z_{L,{\rm lcr}}}(m)
+M_{Z_{L,\nu}}(m)\; .
\end{eqnarray}
Since the swept-up mass is fixed at $M_{\rm sw}=6.5\times 10^4\Msun$
independent of time (cf. \S \ref{mdlgce}), we here denote $M_{\rm sw}$ and
$M_{\rm sh}(m)$ instead of $M_{\rm sw}(m,t)$ and $M_{\rm sh}(m,t)$,
respectively.  We define the ratio of these competing terms as

\begin{equation}
R_{j}(t)=\frac{M_{z_{j},{\rm ISM}}(t)}{{M_{z_{j},{\rm sh}}}(t)}\; ,
\end{equation}
where ${M_{z_j,{\rm sh}}}(t)$ is the IMF-weighted average of the SN-shell term,
${M_{z_j,{\rm sh}}}(m,t)$, over a range of stellar mass $m$.  This ratio gives
a quick measure as to whether a certain element $j$ can be used as an age
indicator of stars born at time $t$ ($ R_{j}(t)>1$), or not ($ R_j(t)<1$).
Figure 6 shows plots of $ R_{j}(t)$ for Fe, B, Be, and $^{6}$Li.

\begin{figure}[p]
\epsfxsize=13cm
\epsfysize=16cm
\epsfbox{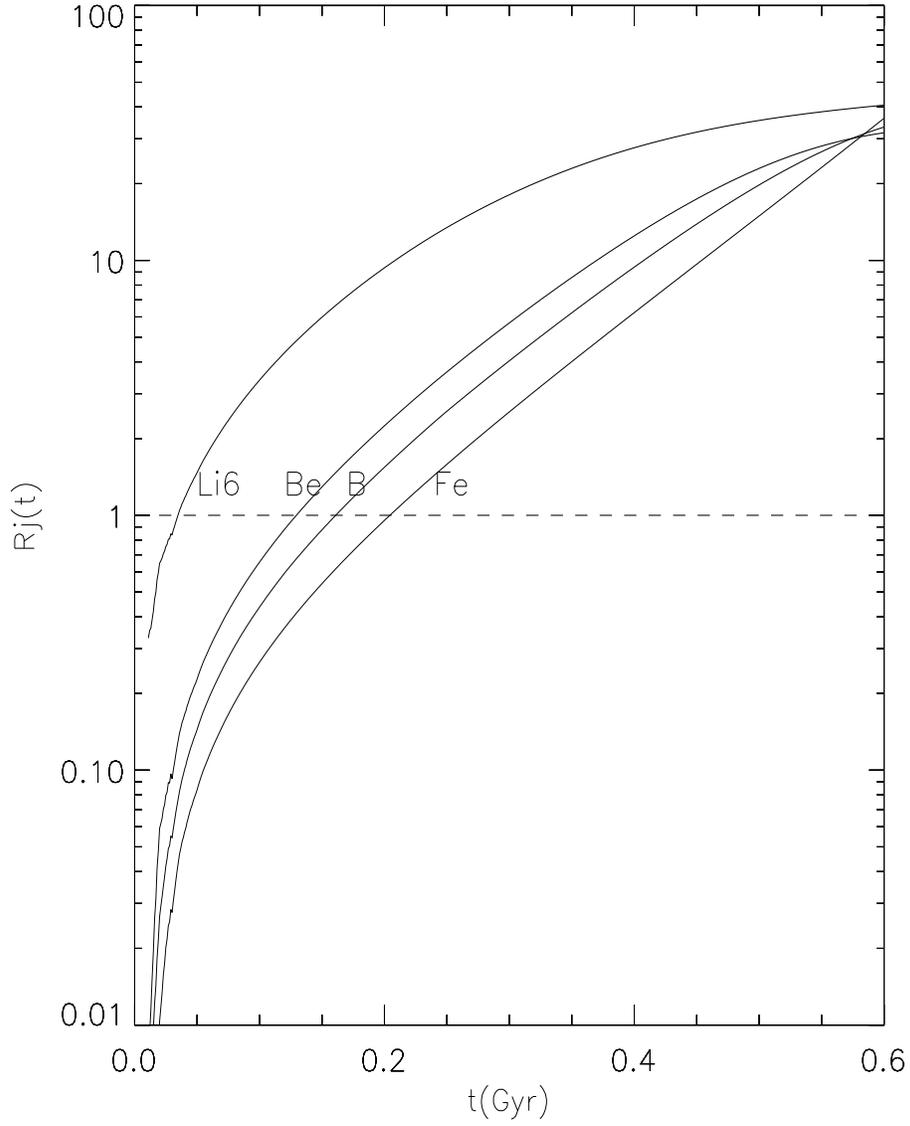}
\caption{ Mass ratio, $R_j(t)$, of the $j$-th element in the SNR
shell, which measures the relative contribution between two different
sources from ISM and stellar ejecta as a function of time
elapsed after the formation of metal-free Pop III stars. Shown are the
cases for $^6$Li, Be, B, and Fe.}
\end{figure}

A certain fraction  of EPs become GCRs which propagate globally, and they will
produce the light elements, not only inside the SNR shells, but also outside of
them.  The light elements outside SNR shells are expected to be distributed
uniformly in the ISM, so that their abundance increases steadily with time.  On
the other hand, the heavy elements produced by SNe II are initially confined in
SNR shells, and remain there until diffusing out as the shells are dissolved.
Thus, the increase of their abundance in the gas is delayed by $\Delta T\sim$3
Myr.  While B originates from both CNO spallation and the $\nu$-process of SNe
II, Be is produced exclusively by CNO spallation, and $^{6}$Li mainly by the
$\alpha + \alpha$ fusion at early epochs.  As mentioned in the beginning of
this subsection, the $\alpha$-particles of BBN origin are distributed
globally, so that $^{6}$Li should also be distributed globally.  As a result,
the value of $R_{\rm ^{6}Li}$ is the largest among the four elements
considered.  As for Be, at early epochs, the process of spallation of CNO
ejected from a SN within its own SNR shell would make a significant
contribution to the Be production in the SN-shell term.  Thus, the rise of
$R_{\rm Be}$ is slower than $R_{\rm ^{6}Li}$.  In addition to this local CNO
spallation, $^{11}$B is also synthesized by the $\nu$-process of SNe II, so
that the rise of $R_{\rm B}$ is slower than $R_{\rm Be}$ and $R_{\rm ^{6}Li}$.

So far we have discussed the relation between time and elemental abundance for
a fixed value of ${E_{\rm lcr}}/{E_{\rm gcr}}=0.1$.  However, different choices
of ${E_{\rm lcr}}/{E_{\rm gcr}}$ would change this relation, because
dominance of local CNO spallation in the SNR shell enhances spatial
inhomogeneity, depending on the mass of the SN progenitor, even for the 
abundance of spallation products.  Figure 7 shows the correlation of $^6$Li and
Be with time for ${E_{\rm lcr}}/{E_{\rm gcr}}= 0,0.1,{\rm and},0.5$.  For the
case of ${E_{\rm lcr}}/{E_{\rm gcr}}= 0$, all the EPs escape from the SNR with
enough energy and become GCRs that propagate globally.  For the case of
${E_{\rm lcr}}/{E_{\rm gcr}}= 0.5$, half of the total acceleration energy is
given to EPs which are eventually thermalized in the same SNR. (This energy
will be used to ``push'' the shock through the adiabatic loss of EPs, or it
will be returned back to internal energy of material in the SNR shell.)

\begin{figure}[p]
\epsfxsize=12cm
\epsfysize=18cm
\epsfbox{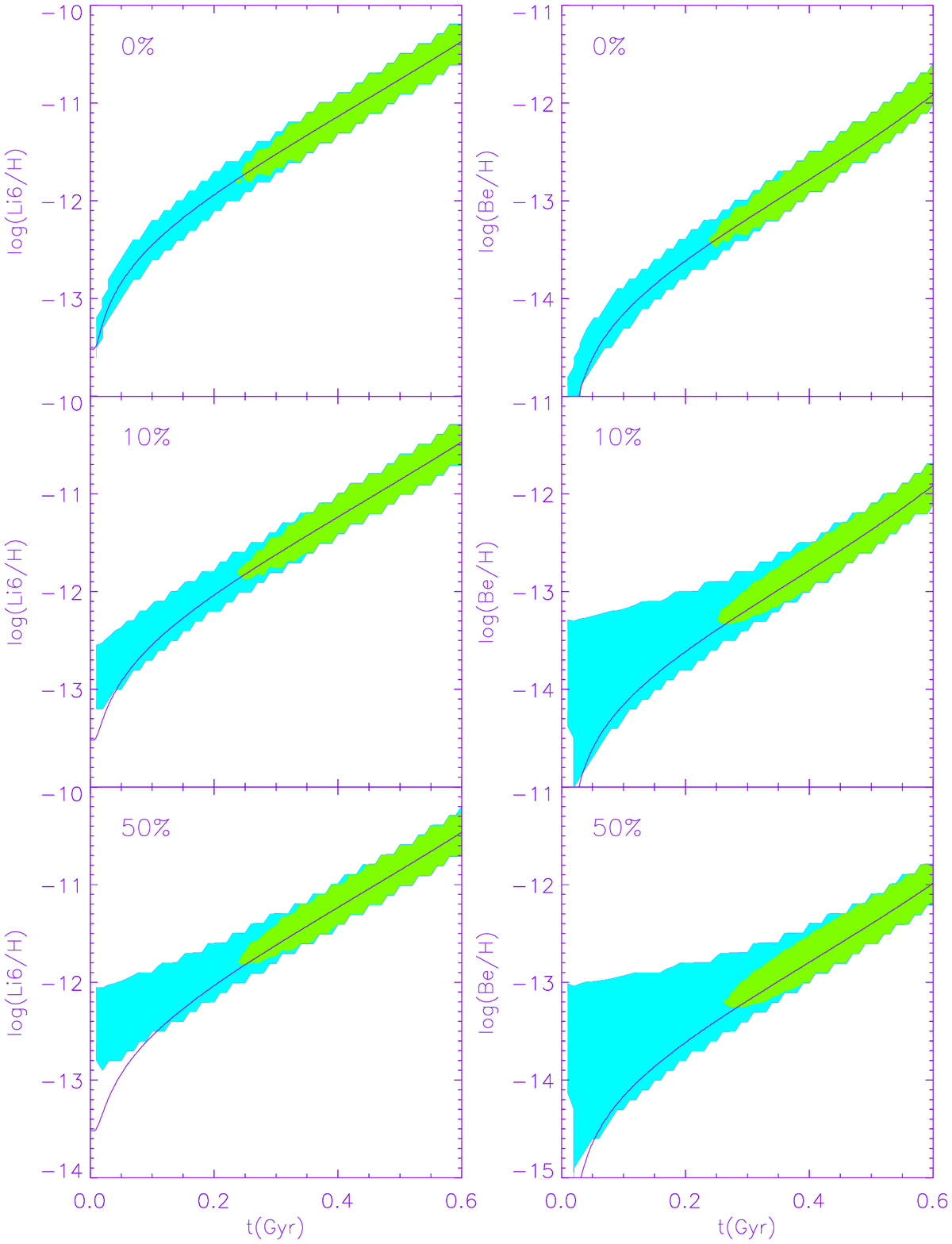}
\caption{ Predicted frequency distribution of long-lived stars in the
$t$-log($X_L$/H) planes for $^6$Li and Be. The same as in Fig.5, but for
various values of the local to global cosmic-ray ratio,
${E_{\rm lcr}}/{E_{\rm gcr}}$. Top two panels show the results for
${E_{\rm lcr}}/{E_{\rm gcr}}=0$, middle panels for ${E_{\rm
cr,l}}/{E_{\rm cr,g}}=0.1$ (same as in Fig. 5), and bottom panels for
${E_{\rm lcr}}/{E_{\rm gcr}}=0.5$. }
\end{figure}

As expected, the correlation between time and elemental abundance is
very good for both $^6$Li and Be for the case of ${E_{\rm lcr}}/{E_{\rm gcr}}=
0$, because these elements are assumed to be produced by global GCRs only.  The
correlation between time and Be becomes worse for larger ${E_{\rm lcr}}/{E_{\rm
gcr}}$ because of the larger contribution of local CNO spallation within the
SNR shell. However, this $t$-Be correlation for the case of ${E_{\rm
lcr}}/{E_{\rm gcr}}= 0.5$ is still better than that of Fe (Fig. 5).  The
abundance of $^6$Li is well-correlated with time even for the case of ${E_{\rm
lcr}}/{E_{\rm gcr}}= 0.5$. This is because $^6$Li is produced by the fusion of
uniformly distributed $\alpha$ particles in the halo.  Figure 7 also shows that
$^6$Li is the best candidate to be used as a cosmic clock, therefore it is
highly desirable to increase the number of stars with measured $^6$Li abundance
in the future.

\section{Discussion}
\label{sec:dscs}

\subsection{Cosmic Ray Energetics and Light-Element Production
--- Comparison with the Superbubble Model}
\label{sec:sbmd}
As discussed in \S \ref{sec:ccpr}, in order to explain the observed $^6$Li and
Be abundances in metal-poor stars, one SN had to provide $E_{\rm gcr}\sim 4
\times 10^{50}$ erg to cosmic rays, even though we adopted a rather softer
spectrum, $\gamma=2.7$, which can produce light elements more efficiently.
This value of GCR energy per SN is higher than that required today, and
is often 
discussed as being problematic (PDa,b; Ramaty \etal 1997, 2000).  The
superbubble model was introduced (Higdon \etal 1998; PDc; Ramaty et al. 2000)
as one possible way to achieve high efficiency of light-element production
while not violating the energy requirements.  According to detailed
calculations by PDc, there are two reasons why superbubbles are suitable sites
for effective production of the light elements.  First, CNO elements are more
efficiently accelerated, because heavy elements ejected by successive
explosions of many (up to $\sim 100$) SNe accumulate inside the superbubble,
and the gas becomes much more metal rich (approaching [Fe/H]$\sim -1$) as
compared to the gas in the outer region ([Fe/H]$\sim -4$).  These energetic CNO
nuclei will be spalled to produce light elements.  Second, the spectrum of EPs
in the superbubble is expected to be $E^{-\alpha}\exp (-E/E_0)$, where $\alpha
\simeq 1\sim 1.5$ and $E_0$ is a few hundred MeV (Bykov, \& Toptygin 1990;
Bykov 1995).  As a result, they more effectively produce light elements than
the so-called momentum spectrum predicted from shock acceleration of the
first-order Fermi process (Blandford \& Ostriker 1978).  An important feature
of this spectrum is an enhancement of EPs in the low-energy region, around
$10-100$ MeV/A, where the cross sections to produce light elements, especially
$^6$Li, have their maximum value (the bottom panel of Fig.1).

\begin{figure}[p]
\epsfxsize=15cm
\epsfysize=12cm
\epsfbox{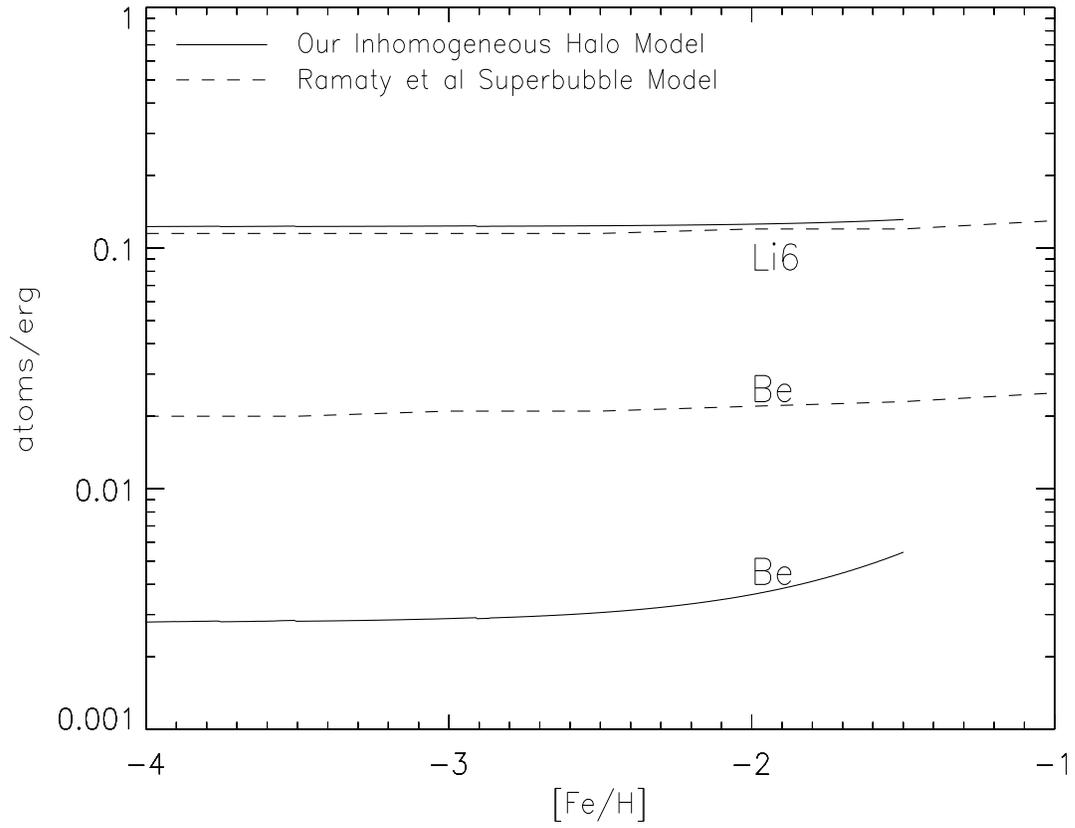}
\caption{Number of Be and $^6$Li atoms produced per unit GCR energy.
Solid lines show the results of our inhomogeneous halo model with
$f_{\rm cr}=0.007$ and $\gamma=2.7$. Dashed lines show the results of
the SN-ejecta-enriched superbubbles by Ramaty \etal (2000).}    
\end{figure}

Here, we would like to compare the superbubble model with our inhomogeneous
halo model with respect to GCR energetics.  In Fig. 8, adopting the spectral
index $\gamma$=2.7, and the parameter $f_{\rm cr}$=0.007, which determines the
primary heavy GCRs (\S \ref{sec:ccpr}), we show the number of Be and $^6$Li
atoms produced per unit GCR energy (atoms/erg) for our model, overlayed with
recent results of the superbubble model by Ramaty et al. (2000).  There are two
primary differences between their superbubble model assumptions and our
individual SN scenario with respect to the nature of the GCRs.  One is the
spectrum of the GCRs: Ramaty et al. (2000) use the spectrum of $q \propto
p^{-\gamma} \exp(-E/{E_0})$, where $p$ is the momentum of a particle and $E_0$
($\sim 10$ GeV/A) is the cut-off energy, calculated on the basis of multiple
shock acceleration in superbubbles (Bykov 1995), while ours is a so-called
momentum spectrum (eq.(\ref{eqn:crfx})) predicted from the Fermi acceleration
mechanism in the shock of individual SNRs. The other difference is the assumed 
chemical composition of GCRs: the CNO abundance by mass adopted in Ramaty \etal
(2000) is almost as high as 0.1, while that adopted in our model lies in the
range 0.005--0.01 during the entire epoch of early Galactic halo (Fig. 4).  It
turns out that their adopted superbubble GCR spectrum yields a similar
production quantity of the light elements as ours, since in the energy region
10--100 MeV/A, in which the light elements are produced the most efficiently,
it is {\it identical} to the momentum spectrum.  This is a result of Ramaty
\etal (2000) adopting a higher cut-off energy ($\sim$ 10 GeV/A) than that in
PDc.  Therefore, both the models yield almost the same number of $^6$Li atoms
per unit GCR energy, much of which is produced by $\alpha$ + $\alpha$ fusion
reactions in addition to the spallation of CNO elements.  On the other hand,
the number of Be atoms produced per unit GCR energy based on our model is as
low as 10\% of the results calculated by Ramaty \etal (2000), because the CNO
abundance in GCRs adoped in our model is lower by about one order of magnitude
than that adopted in their superbubble model.

The efficiency of Be production per unit GCR energy derived from our
model is $\sim$ 10$\%$ of the superbubble model by Ramaty \etal (2000).
Therefore, we adopted a higher energy input, $E_{\rm gcr}=4\times 10^{50}$ erg,
to GCRs per SN, but this value is only a factor of three higher than
that adopted in their model ($E_{\rm gcr}=1.5\times 10^{50}$).  This is 
because the derived Be abundance from their model
is higher by a factor of three at the same [Fe/H] than the results of our
best-fit model to the observations.  Our adopted value of $E_{\rm gcr}=4\times
10^{50}$ erg, though higher than the value of $10^{50}$ erg per SN required to
maintain the energy density of cosmic rays in the current disk (e.g., I
\S 4 in Berezinski$\breve{\rm i}$ \etal 1990), is not unacceptably high,
given that the nature of SNR in the early stage of an evolving halo might be
completely different from those in the current disk, for the following
reasons.  First, metal-poor SNRs suffer little from radiative losses
because of the absence of metals, so they could survive for a longer
period, and should be able to accelerate EPs before losing power and
dissipating.  Second, the SNR shell could keep its identity until the late SNR
phase because merging with other shells could be avoided, owing to the low SN
density in the early Galactic halo---See \S \ref{sec:intr} for order of
magnitude estimates. It is therefore expected that the acceleration of EPs in
the early halo should have been more efficient than for the current disk of the
Galaxy.  The efficiency of acceleration of GCRs is expected to decrease from
the early Galaxy to the present level, as the abundance of heavy elements in
the ISM increases.  Therefore, in more sophisticated models, $E_{\rm gcr}$
should be regarded as a variable with respect to cosmic time, or more
precisely, with respect to the metallicity of the SNR shells.

The abundance pattern of heavy elements in metal-poor stars can distinguish
between the superbubble model and our inhomogeneous halo model, because the
prediction from the superbubble model is completely different from ours.
According to PDc, the superbubble is surrounded by a very metal-poor shell
composed of ISM that is swept up by the expanding bubble. In the bubble, heavy
elements ejected by SNe with various progenitor masses are mixed with the ISM
evaporating from the shell.  Since the timescale of mixing ($\sim$ 1 Myr),
estimated by the size of the bubbles divided by the sound velocity, is smaller
than the timescale of evolution of the bubble ($\sim$ 30 Myr), the chemical
composition of the bubbles are expected to be quite homogeneous.  An important
result of superbubble model by PDc is that stars formed from the material of
the bubble, diffusing into the metal-poor ISM, should exhibit a constant
abundance ratio for any combination of elements, for instance, Be/Fe, and O/Fe,
although their elemental abundances may vary, reflecting the progress of
diffusion.  On the other hand, our model predicts that stars formed in each SNR
shell, originating from various-mass SN progenitors, should exhibit a
remarkable scatter in the abundance ratios of heavy elements.  A realistic
situation may not be so simple, because some stars may have formed in the
superbubble environment, and some may have formed from individual SNR
shells, as recently discussed by Parizot \& Drury (2000).

We suggest that observations of the element Eu is one means by which these two
models could be distinguished.  The synthesized mass of Eu in a SN decreases
with increasing SN progenitor mass, an {\it opposite} trend to the behavior of
the majority of heavy elements, including Fe (Shigeyama \& Tsujimoto 1998).
Therefore, if {\it all} stars at early epochs were formed in individual SNR
shells, they must be distributed along a decreasing line of log(Eu/Fe) as a
function of [Fe/H] (see Fig.3 of Shigeyama \& Tsujimoto 1998).  On the other
hand, if superbubbles are the dominant site of star formation in the early
Galaxy, most stars should have an identical Eu/Fe ratio, and be distributed
along a horizontal line in the [Fe/H]-log(Eu/Fe) plane.  Relevant data
obtained by various observers (Mcwilliam \etal 1995, Ryan \etal 1996, Luck, \&
Bond 1985, Gilroy \etal 1988, and Magain 1989) seem to favor our model.
However, this cannot yet be taken as definitive evidence, because the reported 
data still have quite large errors.  Future observations, especially those with
high accuracy, will hopefully clarify this picture.

\subsection{An AGN in Our Galaxy As Another Accelerator}
\label{sec:agn}
\subsubsection{Activities of Galactic Nucleus}
An active galactic nucleus (AGN), which might have existed in the center of our
Galaxy, could have contributed to the early acceleration of EPs. (Production of
light elements in AGN was studied by Baldwin \etal 1977 and  Crosas \& Weisheit
1996.)  It has been argued from the observed number of QSOs that these objects
were very active at redshift $z\sim2-3$ and stopped their activities by the
present time (Shaver \etal 1996).  Moreover, evolutionary trends of bright QSOs
slightly precedes (Richstone et al. 1998) the trend of cosmic star formation
history (Madau \etal 1996; Madau, Pozzetti, \& Dickinson 1998).  Richstone
\etal (1998) interpreted this fact in such a way that the birth of QSOs was
associated with the formation of the spheroidal components of galaxies, and
pre-dated the active phase of star formation through the evolution of galaxies.
If this scenario is correct, the total energy of a galaxy was supplied first by
an AGN, then gradually by SNe. It is naturally supposed, therefore, that most
EPs at early epochs were accelerated by an AGN/QSO source.  Observations of
stellar motions in the region of the center of the Galaxy show that the mass
density at the very center is quite high ($\ge 2.2\times 10^{12}\Msun\;{\rm
pc^{-3}}$), enough to accommodate a compact object like a massive
blackhole (Genzel \etal 1997), possibly providing evidence of an early AGN in
the Galaxy.

There are two mechanisms which might conceivably supply sufficient energy to
an AGN/QSO.  One is a so-called fuel (accretion-powered) engine, and the other
is a so-called fly-wheel (rotation-powered) engine (Nitta et al. 1991, Nitta
1999).  While the energy supplied by the fuel engine is extracted from the
gravitational energy released from an essentially {\it infinite} amount of the
accreting matter, the energy supplied by the fly-wheel engine originates from
{\it finite} amount of rotation energy of the central blackhole.  In this
scenario, the luminosity of AGN/QSO is expected to evolve as follows:
An AGN/QSO is extremely luminous just after being formed, because either
engine is active, but after a certain period the fly-wheel engine ceases its
activity when the rotation energy is exhausted.  After that, the AGN/QSO
becomes fainter, to a level as low as or less than the Eddington luminosity.
Nitta (1999) discussed the statistical properties of the evolution of AGN/QSOs
using the model of a Kerr-blackhole engine.  According to his work, a typical
AGN with mass of $ 10^8\Msun$ has a lifespan for the fly-wheel engine of
$\sim$ 1 Gyr, which is, interestingly, similar to a typical time scale of the
evolution of the Galactic halo.  The luminosity of such a typical QSO/AGN is
about $10^{46}$ erg s$^{-1}$, which is much larger than the energy input by SN
explosions into the current disk ($\sim 10^{42}$ erg s$^{-1}$).  Even if the SN
rate, or almost equivalently the SFR, was higher by a factor of several
factors of ten in the past (Madau \etal 1998), a typical galactic nuclei could
supply at least by 2 orders of magnitude more energy than the total SNe in
the Galaxy at that time.

Other interesting results concerning activities of the Galactic center
have been obtained from X-ray observations (Koyama et al. 1996), which
detected the existence of a hot plasma containing $\sim 10^{54}$ erg
in the central region.  Judging from the lifetime of the
plasma ($\sim 50000$ yrs), these authors concluded that continuous
energy generation of $10^{41-42}$ erg s$^{-1}$ is required, although a
SN origin is implausible because of various 
observational facts.  They also reported an emission line of iron in the cold
molecular clouds near the Galactic center, possibly due to irradiation from the
center which was brighter in the very recent past ($\sim$300 yrs).  Koyama et
al. (1996) argued from these observations that the Galactic nucleus 
still has intermittent activity, with a time-averaged luminosity $\sim
10^{41-42}$ erg s$^{-1}$, that might result from the the activity of a
fuel engine (accretion-powered) blackhole.  It should be noted that such a
value is quite high, as compared to energy suppliers in our Galaxy today,
being almost comparable to the energy input from the total SN explosions in the
present Galactic disk ($\sim 10^{42}$erg s$^{-1}$).

\subsubsection{Light Element Production by AGN}
If the Galactic nucleus was once very active, with the energy supplied from
the rotation of the central blackhole and/or the accreting matter onto the
blackhole, it is natural to imagine that EPs accelerated in the shock around
the AGN would have enhanced the production of light elements at early epochs.
Adopting typical values of particle velocity $v\sim (0.5-1)\times 10^{10}$
cm$\;$s$^{-1}$ and path length against ionization loss $\Lambda_i 
\sim 1-10$ g$\;$cm$^{-2}$ (Northcliffe \& Schilling 1971) for GCRs with 
$E_{\rm gcr}\sim 10-100$ Mev/A, which most effectively produce the light 
elements, we estimate the lifetime of these GCRs as 
\begin{equation}
\tau_{\rm gcr}=\frac{\Lambda_i}{\rho v}\sim 1-10 {\rm Myr},
\end{equation}
where $\rho$ is the density of ambient gas, having a typical value of 
$\sim 10^{-24}$ g$\;$cm$^{-3}$.  With the diffusion coefficient
$D_{\rm GCR}\sim 10^{29}$ cm$^2$s$^{-1}$ for the Galactic halo 
today (see eq. (\ref{eq:difcr})), these GCRs propagate over the distance
of $1-3$ kpc away from the Galactic center and are expected to enhance
the production of light elements there.

We now consider how these AGN-accelerated GCRs could affect the chemical
evolution of ``clumpy'' clouds which make up the entire Galactic halo at early
epochs.  Studies of the kinematics of stars in the solar neighborhood indicate
that metal-poor halo stars have a variety of orbital eccentricities spanning
from $e\sim 0$ to 1, whereas metal-rich disk stars rotate in almost circular
orbits of $e\sim 0$ in the Galactic disk (Yoshii \& Saio 1979; Norris, Bessell,
\& Pickles 1985; Chiba \& Yoshii 1997).  About 70\% of nearby halo stars with
[Fe/H]$\le-2.2$ have orbits with $e>0.5$ (Chiba \& Beers 2000) which are
accessible to the region of central few kiloparsecs, given that the solar
distance is 8.5 kpc from the Galactic center.  The observed orbital
eccentricities of halo stars more or less reflect the initial motion of clouds
from which such stars were born, because their orbital angular momentum is
preserved, with only rare interactions between the gas and other stars.  Even
if dissipation processes have played a role, initial clouds were dynamically in
a more chaotic state, having more chances to pass through the central Galaxy in
the past.

Accordingly, many halo stars observed in the solar vicinity today were born
from clouds which could pass through the central Galaxy at least once in an
orbital period $T_p$.  Since $T_p$ should be comparable to the dynamical timescale
of the Galaxy $\tau_{\rm dyn,Gal}\sim 10^8$ yr, we obtain the following
inequalities:
\begin{equation}
\label{eq:agnev}
\tau_{\rm halo}(\sim 10^9 {\rm yr}) > T_p(\sim 10^8 {\rm yr})
> \tau_{\rm evol,c}(\sim 2\times 10^7 {\rm yr}) ,
\end{equation}
where $\tau_{\rm halo}$ is a typical duration of the formation of the Galactic 
halo, and $\tau_{\rm evol,c}$ is the typical timescale of cloud evolution,
characterized by the lifetime of $10M_\odot$ stars inducing the formation
of new stars (\S \ref{sec:gcr}).  The first inequality in eq. (\ref{eq:agnev})
indicates that the clouds could have passed near the center of the Galaxy at
least 10 times during the halo phase.  Given that AGN activity lasts for
$\sim 1$ Gyr, the chemical evolution of the clouds is affected by
AGN-accelerated GCRs about 10 times, and the production of light elements is
intermittently enhanced once in every $\sim 10^8$ yr.  Because of the random
and chaotic motion of the clouds, the light elements originating as a result of
AGN activity are distributed over the entire halo, although their production
site is confined in the central few kiloparsecs.  As a result, the
light-element abundance of halo stars would have been increased more rapidly,
compared to the case of an exclusive SNe origin.  The trend of Be and B
abundances with metallicity would be flatter than quadratic, and the trend of
$^6$Li would be even flatter than linear.  These expected consequences are
similar to the model prediction by Yoshii et al. (1997), under the assumption
that the GCR flux was higher in the past, owing to more effective confinement
of GCRs in the early Galaxy.

The second inequality in eq. (\ref{eq:agnev}) indicates that the
AGN-accelerated GCRs affect the production of light elements
intermittently on longer timescales, compared to that of the
stellar generations induced by SNe.  As a result, the trend of
light-element abundance might not be expected to be a 
a simple function of metallicity.  Although precise estimates of 
this extra AGN effect required more detailed modeling of the evolution of 
clumpy clouds in the Galactic halo, it is expected that the light-element 
abundance could still be used as an age indicator, because GCRs originating 
from the AGN propagate as uniformly in the clouds as those from SNe (\S
\ref{sec:gcr}).  
 
\section{Summary}

A model describing the chemical evolution of the light elements has been
constructed by incorporating the inhomogeneous nature in the Galactic halo,
characterized by SN-induced star formation processes (\S \ref{mdlgce}), as well
as the contribution of global GCRs and local EPs in the SNR (\S \ref{sec:gcr}
and \S \ref{sec:leps}).  We have calculated the stellar frequency distribution
in the [Fe/H] {\it vs.} log($X_{L}$/H) plane, which reproduces well the
observed scatter of elemental abundances in halo stars (\S
\ref{sec:cmpobs}).  Inspection of the frequency distribution of stars as a
function of their age (\S \ref{sec:tmr}) indicates that the abundance of light
elements is well-correlated with time.  Based on our model, $^6$Li is the best
cosmic clock, and $^9$Be is the second-best clock.  We have discussed the
allowed combination of various parameters for GCRs/EPs, inferred from
observations of $^6$Li and $^9$Be abundances in metal-poor stars (\S
\ref{sec:ccpr}).  
Although there exist concordant values for these parameters
which account for the observed abundance of light elements in metal-poor stars,
more energy to accelerate EPs is necessary in the early stage of the Galaxy,
compared to that required to maintain GCRs in the current disk.  
We have discussed the energetics of GCRs with respect to the production
of the light elements in comparison with the supperbubble scenario,
which has alternatively been considered as a site of light element
production in metal-poor stars. We further have argued that
observations of the Eu abundance in extremely metal-poor stars are capable of
distinguishing our inhomogeneous halo model from the supperbubble model
(\S \ref{sec:sbmd}). 
We have proposed the hypothesis of AGN activity in our Galaxy as one
source of energy supply at early epochs, and presented a rough model of
the production of the light elements in the Galactic halo arising from AGN
activity (\S \ref{sec:agn}).

We thank Dr. Timothy C. Beers for his valuable comments and considerable 
improvement of presentation in the text. We also thank Drs. Toshitaka
Kajino, Shin-ya Nitta, Masashi Chiba, Toshikazu Shigeyama, and Takuji
Tsujimoto for many fruitful discussions. This work has been supported in part
by a Grant-in-Aid for the Center-of-Excellence research (07CE2002) from the
Ministry of Education, Science, Sports, and Culture of Japan.

\end{document}